\documentclass[iop]{emulateapj}
\usepackage{graphicx}
\usepackage{amsmath}
\usepackage{atbegshi}
\def\hackaltaffiltext#1#2{\AtBeginShipoutNext{\footnotetext[#1]{#2}\stepcounter{footnote}}}

\begin{document}

\title{EXPLORING ANTICORRLELATIONS AND LIGHT ELEMENT VARIATIONS IN NORTHERN GLOBULAR CLUSTERS OBSERVED BY THE APOGEE SURVEY}

\author{
Szabolcs~M{\'e}sz{\'a}ros\altaffilmark{1,2}, 
Sarah~L.~Martell\altaffilmark{3}, 
Matthew~Shetrone\altaffilmark{4}, 
Sara~Lucatello\altaffilmark{5}, 
Nicholas~W.~Troup\altaffilmark{6},
Jo~Bovy\altaffilmark{7},
Katia~Cunha\altaffilmark{8,9}, 
Domingo~A.~Garc\'{\i}a-Hern{\'a}ndez\altaffilmark{10,11},
Jamie~C.~Overbeek\altaffilmark{2},
Carlos~Allende~Prieto\altaffilmark{10,11},
Timothy~C.~Beers\altaffilmark{12},
Peter~M.~Frinchaboy\altaffilmark{13},
Ana~E.~Garc\'{\i}a P\'erez\altaffilmark{6,10,11},
Fred~R.~Hearty\altaffilmark{14}, 
Jon~Holtzman\altaffilmark{15},
Steven~R.~Majewski\altaffilmark{6},
David~L.~Nidever\altaffilmark{16},
Ricardo~P.~Schiavon\altaffilmark{17}, 
Donald~P.~Schneider\altaffilmark{14,18}, 
Jennifer~S.~Sobeck\altaffilmark{6}, 
Verne~V.~Smith\altaffilmark{19}, 
Olga~Zamora\altaffilmark{10,11},
Gail~Zasowski\altaffilmark{20}
}

\altaffiltext{1}{ELTE Gothard Astrophysical Observatory, H-9704 Szombathely, Szent Imre Herceg st. 112, Hungary}
\altaffiltext{2}{Dept. of Astronomy, Indiana University, Bloomington, IN 47405, USA}
\altaffiltext{3}{Dept. of Astrophysics, School of Physics, University of New South Wales, Sydney, NSW 2052, Australia}
\altaffiltext{4}{University of Texas at Austin, McDonald Observatory, Fort Davis, TX 79734, USA}
\altaffiltext{5}{INAF-Osservatorio Astronomico di Padova, vicolo dell’Osservatorio 5, 35122 Padova, Italy}
\altaffiltext{6}{Dept. of Astronomy, University of Virginia, Charlottesville, VA 22904-4325, USA}
\altaffiltext{7}{Institute for Advanced Study, Einstein Drive, Princeton, NJ 08540, USA}
\altaffiltext{8}{University of Arizona, Tucson, AZ 85719, USA} 
\altaffiltext{9}{Observat\'orio Nacional, S\~ao Crist\'ov\~ao, Rio de Janeiro, Brazil}
\altaffiltext{10}{Instituto de Astrof{\'{\i}}sica de Canarias (IAC), E-38200 La Laguna, Tenerife, Spain}
\altaffiltext{11}{Universidad de La Laguna, Departamento de Astrof\'{\i}sica, 38206 La Laguna, Tenerife, Spain}
\altaffiltext{12}{Dept. of Physics and JINA Center for the Evolution of the Elements,
Univ. of Notre Dame, Notre Dame, IN 46556 USA}
\altaffiltext{13}{Texas Christian University, Fort Worth, TX 76129, USA}
\altaffiltext{14}{Dept. of Astronomy and Astrophysics, The Pennsylvania State University, University Park, PA 16802, USA}
\altaffiltext{15}{New Mexico State University, Las Cruces, NM 88003, USA}
\altaffiltext{16}{Dept. of Astronomy, University of Michigan, Ann Arbor, MI 48109, USA} 
\altaffiltext{17}{Astrophysics Research Institute, IC2, Liverpool Science Park,
Liverpool John Moores University, 146 Brownlow Hill, Liverpool, L3 5RF, UK}
\altaffiltext{18}{Institute for Gravitation and the Cosmos, The Pennsylvania State University, University Park, PA 16802, USA}
\hackaltaffiltext{19}{National Optical Astronomy Observatory, Tucson, AZ 85719, USA}
\hackaltaffiltext{20}{Johns Hopkins University, Baltimore, MD, 21218, USA}

\begin{abstract}

We investigate the light-element behavior of red giant stars in Northern globular clusters (GCs) observed by the SDSS-III 
Apache Point Observatory Galactic Evolution Experiment (APOGEE). 
We derive abundances of nine elements (Fe, C, N, O, Mg, Al, Si, Ca, and Ti) for 428 red giant stars in 10 
globular clusters. The intrinsic abundance range relative to measurement errors is examined, 
and the well-known C-N and Mg-Al anticorrelations are explored using an extreme-deconvolution code for the first time 
in a consistent way. We find that 
Mg and Al drive the population membership in most clusters, except in M107 and M71, the two most metal-rich clusters 
in our study, where the grouping is most sensitive to N. We also find a diversity in the abundance 
distributions, with some clusters exhibiting clear abundance bimodalities (for example M3 and M53) while others 
show extended distributions. The spread of Al abundances increases significantly as cluster average 
metallicity decreases as previously found by other works, which we take as evidence that low metallicity, 
intermediate mass AGB polluters were more common in the more metal poor clusters. 
The statistically significant correlation of [Al/Fe] with [Si/Fe] in M15 suggests that 
$^{28}$Si leakage has occurred in this cluster. We also present C, N and O abundances for stars 
cooler than 4500~K and examine the behavior of A(C+N+O) in each cluster as a function of temperature 
and [Al/Fe]. The scatter of A(C+N+O) is close to its estimated uncertainty in all clusters and independent 
on stellar temperature. A(C+N+O) exhibits small correlations and anticorrelations 
with [Al/Fe] in M3 and M13, but we cannot be certain about these relations given the size 
of our abundance uncertainties. Star-to-star variations of $\alpha-$elements (Si, Ca, Ti) 
abundances are comparable to our estimated errors in all clusters.

\end{abstract}

\section{Introduction}

Over the last two decades, the long lasting idea of globular clusters hosting single, 
simple stellar populations has changed dramatically. The classical paradigm 
of GCs being an excellent example of a simple stellar population, defined as a coeval 
and initially chemically homogeneous assembly of stars, has been challenged by observational evidence.
The presence of chemical inhomogeneities, in most cases limited to the light elements 
(the chemical pairs C-N, O-Na, and Mg-Al anti-correlated with each other), have been known for decades and 
recognized to be the signature of high-temperature H-burning. This was initially framed 
within a stellar evolutionary scenario \citep[see, e.g.,][and references therein]{kraft03} given 
that GC abundance work based on high-quality data was limited to bright, evolved giants. 
It was only at the turn of the century that the availability of high-resolution 
spectrographs mounted on 8\,m class telescopes made it possible to carry out studies on the 
compositions of stars down to the main-sequence, which revealed light elements variations 
analogous to those found among giants  \citep{briley96, gratton01, ramirez01}. 
Given that the atmospheres of warm main-sequence stars in large part retain the 
composition of the gas from which they were formed, 
the unavoidable conclusion was that the abundance inhomogeneities are of primordial 
origin.

The most extensive spectroscopic survey of GCs undertaken so far \citep{carretta02, carretta03, carretta01} 
revealed that these inhomogeneities are ubiquitous in Galactic GCs, though 
they do not appear to occur in other star formation environments. However, the
extent of the inhomogeneity varies from 
cluster to cluster, and appears to correlate strongly with the present-day total mass of the GCs, 
and also with metallicity. The improvement in available instrumentation 
and techniques has also led to the discovery 
of a much higher degree of complexity of GC color-magnitude diagrams. 
In fact, while some 
clusters seem to photometrically comply with the simple stellar population paradigm, a growing 
number of them are found to be characterized by multiple main sequences and/or subgiant 
and/or giant branches, \citep[e.g.,]{piotto01, milone01}, which have been associated 
to variations in the content of He and CNO, as well as age spread \citep{dantona02, cassisi01}.

This observational evidence led to a general scenario where GCs host multiple stellar 
populations. These are often assumed to be associated with different stellar generations: 
the ejecta of the slightly older stars, probably mixed with varying amounts of gas from the original 
star forming cloud, creates subsequent younger generation of stars \citep[see e.g.,][]{decressin01, dercole01}, 
though alternative scenarios are also being considered \citep[see, e.g.,][]{bastian01}. It is believed that only a 
fraction of the first-generation of stars can contribute to the internal enrichment. The difference 
in ages among the stellar generations is actually relatively small for the majority of the clusters 
(with a few exceptions such as e.g., $\omega$ Cen or M22) and is confined to a couple of hundreds Myr.

The details of this formation scenario are still far from being understood. 
The origin of the polluting material remains to be established and it has 
obvious bearings on the timescales for the formation of the cluster itself and 
its mass budget.  The observed wide star-to-star variations in C, N, O, Na, and Al found in each Galactic 
GCs, coupled with the uniformity in Fe and Ca (apart from a few notable exceptions) 
provide quite stringent constraints and argue against anything but a minor 
contribution from supernovae \citep{carretta02}. Proposed candidate polluters include 
intermediate mass stars in their asymptotic giant branch (AGB) phase \citep{ventura01}, 
fast rotating massive stars losing mass during their main sequence phase 
\citep{decressin01}, novae \citep{maccarone01} and massive binaries \citep{demink01}.  
These potential contributions obviously operate on different time scales and require a different amount of stellar 
mass in the first-generation. All of the candidates proposed so far fall short of 
reproducing the full variety of observations. Advances in the theoretical modeling of star 
formation and evolution are likely needed to improve our understanding of these issues, 
including the spanning of a larger range of the parameter space (e.g., mass, 
metallicity, and mass loss). However, from an observational point of 
view, the increase in the high-quality 
abundance work available for GCs, both in the sheer number of stars and clusters, as well as
in terms of chemical species considered, is paramount, as it creates a more complete 
picture of the phenomena involved.

The Apache Point Observatory Galactic Evolution Experiment
\citep[APOGEE][]{majewski01} is a three year, near-infrared \citep[15,090 to 16,990 \AA;][]{wil10}, 
high-resolution spectroscopic survey of about 
100,000 red giant stars included as part 
of the 3rd Sloan Digital Sky Survey \citep[SDSS-III][]{eis11}. With a nominal resolving power 
of 22,500, APOGEE is deriving abundances of up to 15 elements for 
nearly 100,000 stars, although fewer elements are generally detected in weak-lined
metal-poor stars. APOGEE is in a unique position among the various Galactic spectroscopic surveys such as Gaia-ESO, 
\citep{gilmore01}, RAVE \citep{ste06}, and GALAH \citep{freeman01}, as it uses the Sloan 2.5m telescope at 
Apache Point Observatory \citep{gunn01}, and thereby has access to the northern hemisphere. 
APOGEE observes a large sample of northern globular clusters, something that makes 
it possible to analyse these clusters 
in a homogeneous way, which has not been done before for these objects. 

The study of GCs with APOGEE plays an important role not just because it has access to many northern GCs. 
Its high-resolution near-IR spectra allow the simultaneous determination of many elemental abundances generally 
not available in optical spectroscopic work of GC stars. C and N, which are elements 
heavily affected by the pollution phenomenon in GCs, are often not included
in studies of metal poor stars because the strongest features (CH and CN) lie in the 
near-UV, far from the optical lines of Na, Mg and Al, and thus multiple 
detectors or setups are required to obtain both sets.    
In addition, because these studies usually focus on fairly red stars longer 
exposure times are required to acquire sufficient SNR to analyse the near-UV
features.

The spectra used in this paper are publicly available as part of the tenth data release \citep[DR10,][]{ahn01} 
of SDSS-III. The initial set of stars selected were the same 
used by \citet{meszaros02} to check the accuracy and precision of APOGEE parameters published in DR10. 
However, instead of using the automatic ASPCAP pipeline, we will make use of photometry and theoretical isochrones 
to constrain the effective temperature (T$_{\rm eff}$) and surface gravity $\log~g$, and use an 
independent semi-automated method for elemental abundance determination for 10 northern GCs. Some of 
these clusters are well studied, such as M3, M13, M92, M15, while others have been poorly studied (NGC 5466), 
or been only recently discussed in the literature, such as M2 \citep{yong06}.

\section{Abundance Analysis}

\subsection{Target Selection}

Table 1 lists the globular clusters APOGEE observed in its first year, along with the adopted
[Fe/H], E($B-V$), and ages from the literature.  Targets were 
selected as cluster members if 1) there is published abundance 
information on the star as a cluster member, 2) the star is a radial velocity 
member, or 3) if it has a probability $>$50\%  of being a cluster 
member based on their proper motion adopted from the literature. After this initial selection, we checked 
the position of stars in the T$_{\rm eff} - \log~g$ diagram based on APOGEE observations and deleted those 
that were not red giant branch (RGB) stars. Stars that have metallicity 0.3~dex (typically 3~$\sigma$ scatter) larger or smaller than the 
cluster average also need to be deleted, but this last step resulted in no rejections. The cluster target 
selection process is described in more detail in \citet{zasowski01}. 
The final sample consists of 428 stars from 10 globular clusters. 
High S/N spectra are essential to determine abundances from atomic and 
molecular features, thus all selected targets have at least S/N=70 as 
determined by \citet{meszaros02}.

\begin{deluxetable}{llrrc}
\tabletypesize{\scriptsize}
\tablecaption{Properties of Clusters from the Literature}
\tablewidth{0pt}
\tablehead{
\colhead{ID} & \colhead{Name} & \colhead{N\tablenotemark{a}} &
\colhead{[Fe/H]\tablenotemark{b}} & 
\colhead{E(B$-$V)Ref.\tablenotemark{c}} }
\startdata
NGC 7078	& M15		& 23 & -2.37$\pm$0.02 &  0.10  \\
NGC 6341	& M92		& 47 & -2.31$\pm$0.05 &  0.02 \\
NGC 5024	& M53		& 16 & -2.10$\pm$0.09 &  0.02 \\
NGC 5466	& 			& 8 & -1.98$\pm$0.09 &  0.00   \\
NGC 6205	& M13		& 81 & -1.53$\pm$0.04 &  0.02 \\
NGC 7089	& M2		& 18 & -1.65$\pm$0.07 &  0.06   \\
NGC 5272	& M3		& 59 & -1.50$\pm$0.05 &  0.01 \\
NGC 5904	& M5		& 122 & -1.29$\pm$0.02 &  0.03  \\
NGC 6171	& M107  	& 42 & -1.02$\pm$0.02 &  0.33  \\
NGC 6838	& M71		& 12 & -0.78$\pm$0.02 &  0.25  \\
\enddata
\tablenotetext{a}{N is the number of stars observed in each cluster.}
\tablenotetext{b}{[Fe/H] references: \citet{harris01}, clusters are listed in order 
of the average cluster metallicity determined in this paper.}
\tablenotetext{c}{E(B$-$V) references: (1) \citet{harris01}.}
\end{deluxetable}

\begin{deluxetable*}{lrrrrrrrrrrrrr}
\tabletypesize{\scriptsize}
\tablecaption{Properties of Stars Analyzed}
\tablewidth{0pt}
\tablehead{
\colhead{2MASS ID} & 
\colhead{Cluster} & 
\colhead{v$_{\rm helio}$} &
\colhead{T$_{\rm eff}$} & 
\colhead{$\log~g$} & 
\colhead{[Fe/H]} & 
\colhead{[C/Fe]} &  
\colhead{[N/Fe]} & 
\colhead{[O/Fe]} &
\colhead{[Mg/Fe]} & 
\colhead{[Al/Fe]} & 
\colhead{[Si/Fe]} & 
\colhead{[Ca/Fe]} & 
\colhead{[Ti/Fe]} 
}
\startdata
2M21301565+1208229 & M15 & -104.5 & 4836 & 1.56 & -2.12 & \nodata & \nodata & \nodata & 0.16 & -0.06 & 0.35 & 0.19 & \nodata \\
2M21301606+1213342 & M15 & -108.3 & 4870 & 1.64 & -2.31 & \nodata & \nodata & \nodata & 0.10 & 0.57 & 0.46 & 0.53 & \nodata \\
2M21304412+1211226 & M15 & -102.7 & 4715 & 1.28 & -2.12 & \nodata & \nodata & \nodata & -0.45 &  0.63 & 0.60 & 0.35 & \nodata \\   
2M21290843+1209118 & M15 & -106.0 & 4607 & 1.03 & -2.07 & \nodata & \nodata & \nodata & -0.11 &  0.75 & 0.41 & \nodata & \nodata \\
2M21294979+1211058 & M15 & -107.6 & 4375 & 0.56 & -2.31 & -0.44 & 0.95 & 0.44 & 0.17 & 0.64 & 0.44 & 0.06 & \nodata \\
\enddata
\tablecomments{
This table is available in its entirety in machine-readable form in the online journal. A portion 
is shown here for guidance regarding its form and content. }
\end{deluxetable*}

\subsection{Atmospheric Parameters}

Abundances presented in this paper are defined for each individual element X heavier than helium as

\begin{equation}
[X/H] = \log_{10}(n_X/n_H)_{\rm star} - \log_{10}(n_X/n_H)_\odot 
\end{equation}

where $n_X$ and $n_H$ are respectively the number of atoms of
element X and hydrogen, per unit volume in the stellar photosphere.

To derive abundances from stellar spectra, we first have to estimate four main 
atmospheric parameters: T$_{\rm eff}$, $\log~g$ , microturbulent velocity, 
and overall metallicity ([Fe/H]). In the following sub-sections we present our
methodology for determining these parameters and our reasons for not using the values available for each star from 
the APOGEE Stellar Parameters and Chemical Abundances Pipeline \citep[ASPCAP;][]{perez01} in DR10. 
In order to evaluate the accuracy and precision of the ASPCAP parameters, \citet{meszaros02}  
carried out a careful comparison with literature values using 559 stars in 
20 open and globular clusters. 

These clusters were chosen to cover most of the 
parameter range of stars APOGEE is expected to observe. 
\citet{meszaros02} provided a detailed explanation of the accuracy and precision of these parameters, 
and also derived empirical calibrations for T$_{\rm eff}$, $\log~g$, and [M/H] using literature data.   
In the sections below we will briefly review these calibrations along with their limitations.

\subsection{The Effective Temperature}

We adopt a photometric effective 
temperatures calculated from the $J-K_{s}$ colors using the equations of \citet{gonzalez01}. 
Their calibration was chosen because of its proximity of only 30$-$40~K 
to the absolute temperature scale.
De-reddened $J-K_{s}$ were calculated 
the same way as by \citet{meszaros02}, from $E(B-V)$, listed in Table 1 for each cluster, 
using $E(J-K_{s}) = 0.46 \cdot E(B-V)$. 

The ASPCAP DR10 effective temperatures were compared with photometric ones using 
calibrations by \citet{gonzalez01} based on 2MASS $J-K_{s}$ colors \citep{struskie01}. 
\citet{meszaros02} 
found that small systematic differences, in the range of 100$-$200K, 
are present between the ASPCAP and photometric temperatures. 
The ASPCAP DR10 T$_{\rm eff}$ were also compared to literature spectroscopic temperatures, 
and the average of these differences were found to be negligible.  
The corrected ASPCAP DR10 temperatures were calculated between 
3500~K and 5500~K using a calibration relation derived from the comparison with the \citet{gonzalez01} 
scale. ASPCAP DR10 raw temperatures above 5000~K in metal-poor stars showed significant, 
300$-$500~K offsets compared to photometry, and are thus believed to be not accurate enough for abundance analysis. 
This issue is mostly limited to metal poor stars, and does not affect stars at [Fe/H]$> -$1, 
where the vast majority of APOGEE targets are.

The adoption of a purely photometric temperature scale enables us to be somewhat 
independent of ASPCAP (while still using the same spectra), which gives important comparison 
data for future ASPCAP validation. 
Besides providing an independent comparison dataset for APOGEE, the photometric temperatures allowed us to 
include stars in the sample that are hotter than 5000~K. Because of these reasons, the final results 
presented in this paper are based on the photometric temperatures, and we only use the ASPCAP DR10 raw 
temperatures to estimate our errors related to the atmospheric parameters.

\subsection{The Metallicity}

The APOGEE DR10 release contains 
metallicities derived by ASPCAP for all stars analysed, 
thus providing an alternative scale to manually derived metallicities. 
That metallicity, [M/H], tracks all metals relative 
to the Sun, and gives the overall metallicity of the stars because it was derived by 
fitting the entire wavelength region covered by the APOGEE spectrograph. This is different 
from most literature publications that use Fe lines to track metallicity in a stellar atmosphere. 
We use [Fe/H] in this paper whenever we refer to metallicity presented here, because we use Fe I lines 
to measure it. For the most part, one can treat values of [M/H] as if they were [Fe/H].
When ASPCAP metallicity was compared with individual values from high-resolution 
observations from the literature, a difference of 0.1~dex is found below [M/H]=$-$1, and 
this discrepancy increased with decreasing metallicity reaching 0.2$-$0.3 dex around 
[M/H]=$-$2 \citep{meszaros02}. The calibrated DR10 metallicities map well onto [Fe/H], 
because the calibration process uses [Fe/H] values from the optical.

Because an alternative metallicity based on only iron lines did not 
exist for APOGEE in DR10, and because of these small zero point offsets, we decided to derive 
our own [Fe/H] based on Fe I lines found in the H band, instead 
of using the published APOGEE DR10 [M/H] values. We assumed that all stars have the 
literature cluster mean metallicity before starting the calculations, after 
which small wavelength windows 
around the Fe lines listed in Table 3 were used to revise the individual 
star metallicities.

\subsection{The Surface Gravity}

In this study we adopt gravities from stellar evolution calculations. Following \citet{meszaros02}, we derive gravities for
our sample using isochrones from the Padova group \citep{bertelli01, bertelli02}. 
The cluster metallicities collected from the 
literature used in the isochrones are listed in Table 1, while the ages of all clusters were chosen to be 10 Gyr. 
The final set of temperatures and gravities corresponding to them from the isochrones are listed in Table 2.

The ASPCAP DR10 surface gravities were compared to both isochrones and Kepler \citep{bo10} asteroseismic
targets observed by APOGEE \citep{pinn01} to estimate their accuracy and precision. 
An average difference of 0.3~dex was found at solar metallicity in both cases, but this  
increased to almost 1~dex for very metal poor stars. The asteroseismic gravities are believed to have 
errors in the range of 
0.01$-$0.05~dex, thus far superior to spectroscopic measurements, but they are only available for 
metal rich stars with [M/H]$ > -$1.0. The discrepancy between spectroscopic and 
asteroseismic surface gravities is a topic of ongoing investigation \citep[e.g.,][]{pinn01, epstein01}. 
The final calibration from \citet{meszaros02} 
combined surface gravities derived from isochrones below [M/H]$ < -$1.0 with 
the asteroseismic dataset for high metallicities.  
Values after the calibration still show some small ($<$0.1~dex) offsets and large scatter 
around the isochrones for the globular clusters.
Thus, in order to minimize errors in abundances 
related to the uncertainty in the ASPCAP corrected spectroscopic gravities, we decided to 
use the pure isochrone gravities in this study. 

\subsection{The microturbulent velocity}

A relation between microturbulent velocity and surface gravity, 
${\rm v}_{\rm micro} = 2.24 - 0.3 \times \log g$, 
was used for ASPCAP analyses in DR10 \citep{meszaros02}. For consistency with 
ASPCAP, and for easier future comparisons with APOGEE results, we adopted 
that equation in this work.

\section{\textit{Autosynth}}

The program called \textit{autosynth} was developed especially for this project to 
simplify the large amount of synthesis required for abundance determination. 
The program takes atmospheric parameters as input and carries out spectral synthesis to derive 
elemental abundances. The core of \textit{autosynth} is MOOG2013\footnotetext{http://www.as.utexas.edu/~chris/moog.html}
\citep{sneden07}, which does the spectrum synthesis, while 
\textit{autosynth} compares the synthetic spectrum with the observed spectrum and determines the best 
abundances with $\chi^2$ minimization in wavelength windows specified by the user. The program can 
read the MOOG formatted ATLAS and MARCS model atmospheres, and can also convert the original 
ATLAS and MARCS formats into a MOOG compatible format. 

The line list adopted for this study 
includes both atomic and molecular species. It is an updated version of what was used for the DR10 results, 
version m201312160900 \citep[also used for DR12, APOGEE's next public data release;][]{alam01, hol01},
and includes atomic and molecular species. The molecular line
list is a compilation of literature sources including transitions of 
CO, OH, CN, C$_2$, H$_2$, and SiH. All molecular data are 
adopted without change with the exception of a few obvious typographical
corrections.  The atomic line list was compiled from a
number of literature sources and includes theoretical, astrophysical and
laboratory oscillator strength values. These literature line positions, oscillator 
strengths, and damping values were allowed to vary in order to fit to the solar 
spectrum and the spectrum of Arcturus, thus generating a tuned astrophysical line list. 
The solution is weighted such that the solar solution has twice the weight as the 
Arcturus solution to properly consider the fact that the abundance ratios in Arcturus 
are more poorly understood than those of the Sun. 
The code used for this process was based on
the LTE spectral synthesis code MOOG \citep{sneden07} but adapted to our unique
needs. For lines with laboratory oscillator strengths, we did not allow
the astrophysical gf value to vary beyond twice the error quoted by the source.
A more detailed description of this process and the line list can be found 
in \citet{shetrone03}.

The choice for the local continuum set can greatly affect the derived abundances, thus we needed a reliable automated 
way to determine the continuum placement. This was done with a separate $\chi^2$ minimization from the one that 
was used for the abundance determination. Continuum normalized 
observation points around 1 are multiplied by a factor between 0.7 and 1.1 with 0.001 steps for 
each synthesis emulating slightly different choices for the location of the local continuum. Multiplication is necessary 
because it preserves the original spectrum. The $\chi^2$ near the continuum is calculated and compared to the 
continuum of the observation and minimized separately for every abundance step. 
The $\chi^2$ calculation for the abundances determination happens between certain flux ranges 
(usually between 0.3 and 1.1) using the continuum placement determined in the previous step.

\subsection{Individual Abundances}

\begin{figure*}[!ht]
\centering
\includegraphics[width=5in,angle=0]{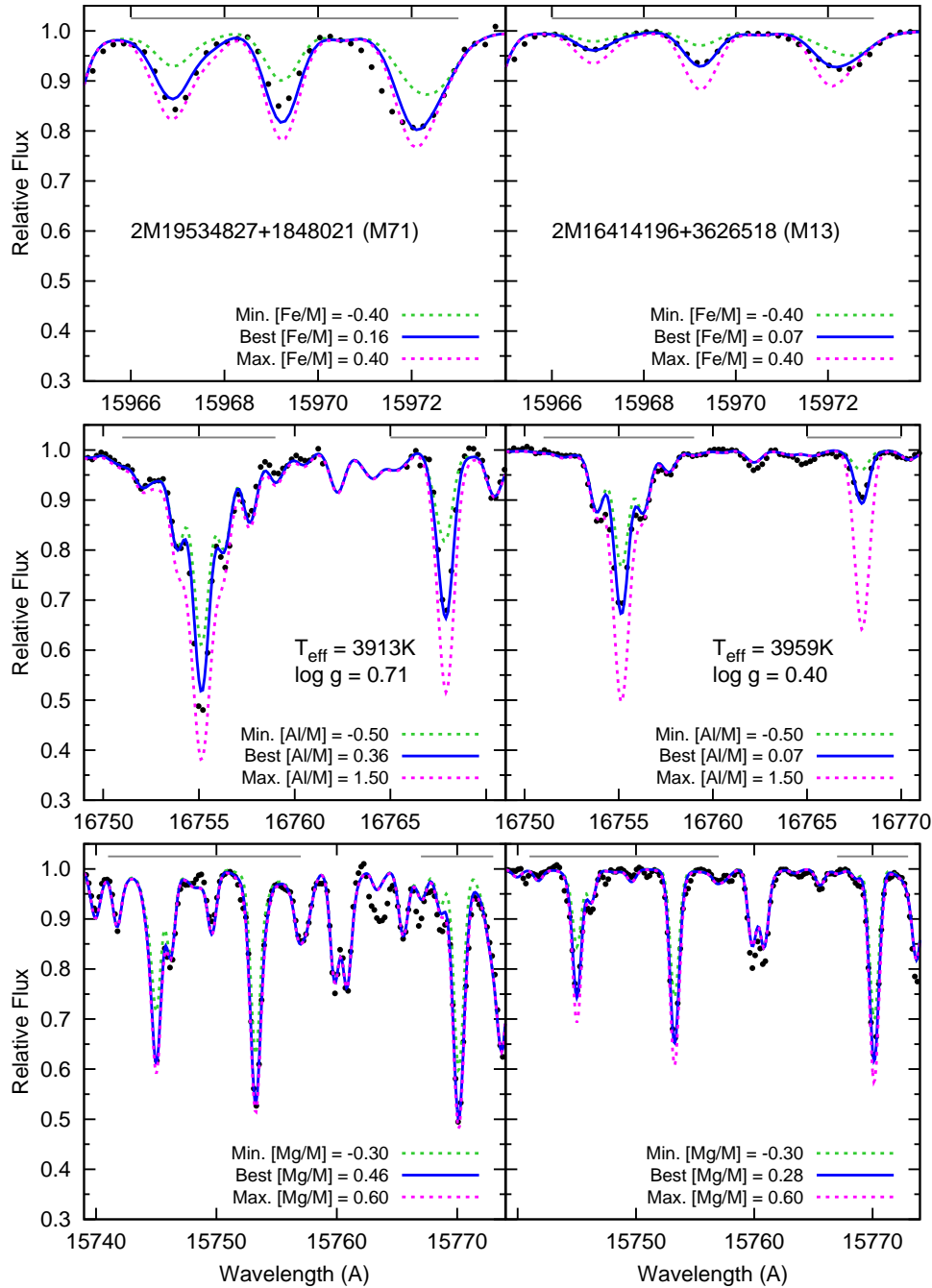}
\caption{Example spectra and the fitted synthesis of Fe, Al, and Mg lines for two stars from M71 and M13. Abundances were 
fitted between the labelled minimum and maximum values using a step of 0.01~dex. The printed best fitted 
abundance values might not be the same as in Table 2, because the table contains averaged values, not individual fits.
}
\label{fig:spectra1}
\end{figure*}

\begin{figure*}[!ht]
\centering
\includegraphics[width=5in,angle=0]{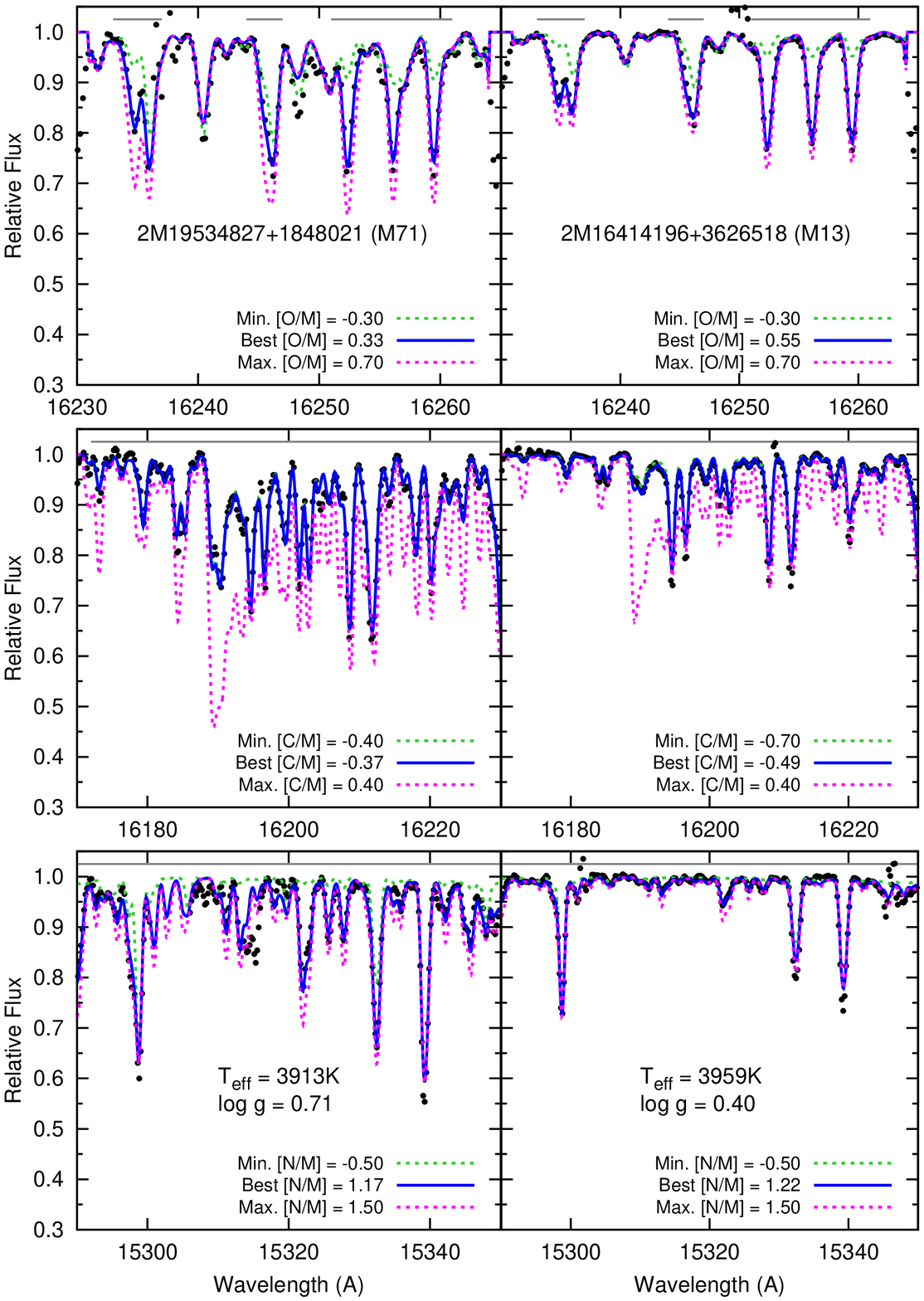}
\caption{Example spectra and the fitted synthesis of OH, CO, and CN lines for two stars from M71 and M13. For more 
explanation see caption of Figure~\ref{fig:spectra1} and Section 3.1.
}
\label{fig:spectra2}
\end{figure*}

The individual abundances were determined using the 1D Local Thermodynamic Equilibrium (LTE) model 
atmospheres calculated with ATLAS9 \citep{kurucz05}. The model atmospheres were generated using 
solar reference abundances from \citet{asplund01}, the same way as the main APOGEE model 
atmosphere database was generated \citep{meszaros01}. Because the overall metallicity of 
these clusters were well known from the literature, initially we calculated atmospheres using the 
average literature metallicity for each cluster adopting the photometric effective temperatures and 
isochrone gravities. These initial model atmospheres were later revised to have consistency 
with the synthesis. 

The windows used to derive the individual abundances were determined based on the analysis of FTS 
stars in the H$-$band using the APOGEE line list by \citet{smith01}. In the case of Fe we measured [Fe/M], 
relative to the literature cluster metallicity for each line. The abundance of Na is very important 
in discussing the spread of O in GCs, and two Na lines are available in the APOGEE spectral band. However, these two Na lines are 
weak even at solar metallicities. We carried out a number of tests attempting to derive Na abundances, but we 
found that the two Na lines become very weak around [Fe/H]=$-$0.5, and non-detectable below about $-0.7$, thus 
we were not able to determine Na abundances for any of our targets. 
The list of wavelength regions used in our analysis, and the solar reference values for each element is listed in 
Table 3. Figures \ref{fig:spectra1} and \ref{fig:spectra2} show examples of observed Fe, Mg, Al, OH, CO, and CN 
line profiles and their fitted synthesis for one star from M71 and M13. The wavelength regions shown in these figures 
are only a fraction of what has been used from Table 3. 

CN lines spread over most of the H$-$band, hence it is important to calculate the CNO abundances 
before the atomic ones. It is also important to use self consistent model atmospheres, because stars 
in globular clusters exhibit low carbon and high $\alpha-$content, which significantly alters the 
structure of the atmosphere compared to a solar scaled one \citep{meszaros01}. Taking into account all this, 
we developed the following procedure to produce the final abundances for each star:

\begin{deluxetable}{lrc}
\tabletypesize{\scriptsize}
\tablecaption{Wavelength Regions}
\tablewidth{0pt}
\tablehead{
\colhead{Element} & \colhead{log(N)\tablenotemark{a}} &
\colhead{Wavelength (\AA)\tablenotemark{b}}  }
\startdata
Fe	& 7.45 & 15210-15213.5; 15397-15401; 15651-15654 \\
	&  & 	15966-15973; 16044-16048; 16156-16160 \\
	&  & 	16168-16171 \\
C	& 8.39 & 15572-15606; 15772-15791; 15980-16037 \\
	&  & 16172-16248; 16617-16677; 16839-16870 \\
N	& 7.78 & 15240-15417 \\
O	& 8.66 & 15267-15272; 15281-15288; 15372-15380 \\
	&  &  15386-15390; 15394-15397; 15404-15414 \\
	&  &  	15499-15502; 15508-15511; 15539-15542 \\
	&  &  15561-15566; 15569-15574; 15887-15904 \\
	&  &  16188-16198; 16207-16213; 16233-16237 \\
	&  & 16244-16247; 16251-16261; 16300-16305 \\
	&  &  16314-16319; 16707-16714; 16718-16720 \\
	&  &  16731-16735; 16888-16892; 16898-16912 \\
Mg	& 7.53 & 15741-15757; 15767-15773 \\
Al	& 6.37 & 16720-16727; 16751-16759; 16765-16770 \\
Si	& 7.51 & 15962-15966; 16062-16066; 16097-16101 \\
    &  & 16218-16223; 16683-16687; 16830-16834 \\
Ca	& 6.31 & 16139-16143; 16153.5-16164 \\
Ti	& 4.90 & 15546.5-15549.5; 15718-15721.5 \\
\enddata
\tablenotetext{a}{The Solar reference abundances are from \citet{asplund01}.}
\tablenotetext{b}{Vacuum wavelength.}
\end{deluxetable}

\begin{enumerate}
\item{A model atmosphere is generated using literature cluster average metallicities, the photometric temperature 
and an isochrone gravity. Because all of our targets are RGB stars, we choose [C/Fe] $= -0.5$, [O/Fe] $= 0.3$, 
and [N/Fe] $= 0.5$ dex for this initial model. }
\item{Individual Fe I lines are fit with \textit{autosynth}, and an average [Fe/H] is calculated for each star.}
\item{A new model atmosphere is calculated using this new [Fe/H] value, but still using the starting CNO abundances.}
\item{We set the abundances of C, N and O before the remaining elements, 
because they can have a significant effect on the atmospheric structure in 
cool stars. Since molecular features generally disappear from metal-poor 
spectra above 4500~K, we divide our stars into two temperature groups. For 
the stars cooler than 4500~K, we first determine [O/Fe] using OH lines, 
then create a new model atmosphere with [$\alpha$/Fe] equal to [O/Fe]. We 
then determine C and O abundances from CO lines, then recreate the model 
atmosphere again with these new [C/Fe] and [O/Fe] abundances. Finally, we 
derive N abundance using CN lines. For stars hotter than 4500~K, we leave 
the C, N, and O abundances at their inital values.  }
\item{The abundances of the remaining elements (Mg, Al, Si, Ca and Ti) are 
determined with \textit{autosynth}, using the stellar parameters, metallicities, 
and C, N and O abundances previously determined.}
\end{enumerate}

For each element, we average together the abundance results from the different wavelength regions to obtain final values. 
Although the size of each region is different, we did not find it necessary to use 
weights based on their ranges or line strengths, because that approach did not produce abundances significantly 
different from a straightforward average. Data reduction errors or missing data affected some 
of these regions, resulting in erroneous fits, and because of this we carefully examined each fit by eye. 
These wavelength regions were not included when constructing the final average abundances. The final abundance values are 
listed in Table 2.

\subsection{Uncertainty Calculations}

\subsubsection{Systematic Uncertainties}

\begin{deluxetable}{lccccccccc}
\tabletypesize{\scriptsize}
\tablecaption{Estimated Abundance Uncertainties}
\tablewidth{0pt}
\tablehead{
\colhead{Cluster} &
\colhead{Fe} &
\colhead{C} &
\colhead{N} &
\colhead{O} &
\colhead{Mg} &
\colhead{Al} &
\colhead{Si} &
\colhead{Ca} &
\colhead{Ti} 
}
\startdata
\cutinhead{Systematic uncertainties from atmospheric parameters}
$\Delta$T$_{\rm eff}$ & 0.08 & 0.12 & 0.16 & 0.18 & 0.04 & 0.07 & 0.05 & 0.06 & 0.15 \\
=150K & & &  &  & & & &  & \\
$\Delta$T$_{\rm eff}$ & 0.05 & 0.08 & 0.11 & 0.12 & 0.03 & 0.05 & 0.03 & 0.04 & 0.10 \\
=100K & &  & &  & &  & &  &\\
\cutinhead{Random uncertainties from \textit{autosynth}}
M15		& 0.11 & 0.21 & 0.30 & 0.06 	& 0.08 & 0.16 & 0.13 & 0.19 & 0.26 \\
M92		& 0.10 & 0.18 & 0.30 & 0.05 	& 0.07 & 0.12 & 0.12 & 0.22 & 0.17 \\
M53		& 0.06 & 0.16 & 0.12 & 0.05 	& 0.06 & 0.14 & 0.11 & 0.11 & 0.18 \\
N5466	& 0.06 & 0.18 & 0.15 & 0.03 	& 0.07 & 0.03 & 0.07 & 0.14 & 0.17 \\
M13		& 0.05 & 0.12 & 0.06 & 0.05 	& 0.05 & 0.12 & 0.07 & 0.06 & 0.12 \\
M2		& 0.04 & 0.11 & 0.06 & 0.04 	& 0.07 & 0.10 & 0.09 & 0.05 & 0.09 \\
M3		& 0.04 & 0.12 & 0.08 & 0.05 	& 0.03 & 0.10 & 0.08 & 0.05 & 0.12 \\
M5		& 0.04 & 0.08 & 0.05 & 0.05 	& 0.04 & 0.06 & 0.07 & 0.04 & 0.09 \\
M107	& 0.04 & 0.04 & 0.03 & 0.05 	& 0.04 & 0.08 & 0.07 & 0.04 & 0.10 \\
M71		& 0.02 & 0.03 & 0.04 & 0.04 	& 0.04 & 0.05 & 0.07 & 0.06 & 0.07 \\
\cutinhead{Final combined uncertainties}
M15		& 0.12 & 0.22 & 0.32 & 0.13 	& 0.09 & 0.17 & 0.13 & 0.19 & 0.28 \\
M92		& 0.11 & 0.20 & 0.32 & 0.13 	& 0.09 & 0.13 & 0.12 & 0.22 & 0.20 \\
M53		& 0.08 & 0.18 & 0.16 & 0.13 	& 0.07 & 0.15 & 0.11 & 0.12 & 0.21 \\
N5466	& 0.08 & 0.20 & 0.19 & 0.12 	& 0.08 & 0.06 & 0.08 & 0.15 & 0.20 \\
M13		& 0.07 & 0.14 & 0.13 & 0.13 	& 0.06 & 0.13 & 0.08 & 0.07 & 0.16 \\
M2		& 0.06 & 0.14 & 0.13 & 0.13 	& 0.08 & 0.11 & 0.09 & 0.06 & 0.13 \\
M3		& 0.06 & 0.14 & 0.14 & 0.13 	& 0.04 & 0.11 & 0.09 & 0.06 & 0.16 \\
M5		& 0.06 & 0.11 & 0.12 & 0.13 	& 0.05 & 0.08 & 0.08 & 0.06 & 0.13 \\
M107	& 0.06 & 0.09 & 0.11 & 0.13 	& 0.05 & 0.09 & 0.08 & 0.06 & 0.14 \\
M71		& 0.05 & 0.09 & 0.12 & 0.13 	& 0.05 & 0.07 & 0.08 & 0.07 & 0.12 \\
\enddata
\tablecomments{Top panel: Systematic uncertainty estimates from changes 
in T$_{\rm eff}$, $\log~g$, and $v_{micro}$. 
Middle panel: Average random uncertainties reported by \textit{autosynth}.
Bottom panel: The final uncertainties are the sum of uncertainties in quadrature from the middle 
panel and uncertainties for $\Delta$T$_{\rm eff}$=100K, $\Delta~\log~g$=0.3, and 
$\Delta~v_{micro}$=0.1~km/s.}
\end{deluxetable}

The uncertainty in the atmospheric parameters strongly affects the final 
abundances derived from some of the spectral features we consider.  
To test the sensitivity of abundances due to changes in the atmospheric 
parameters we used the results from the ASPCAP raw temperature scale.  
The same exact steps described in the 
previous section were followed, but instead of adopting the photometric 
temperature scale we adopt the ASPCAP DR10 raw temperature scale, which results in 
new surface gravities and microturbulent velocities. This way we could track 
systematics uncertainties sensitive to these parameters as well.

\begin{figure}[!ht]
\centering
\includegraphics[width=3.5in,angle=0]{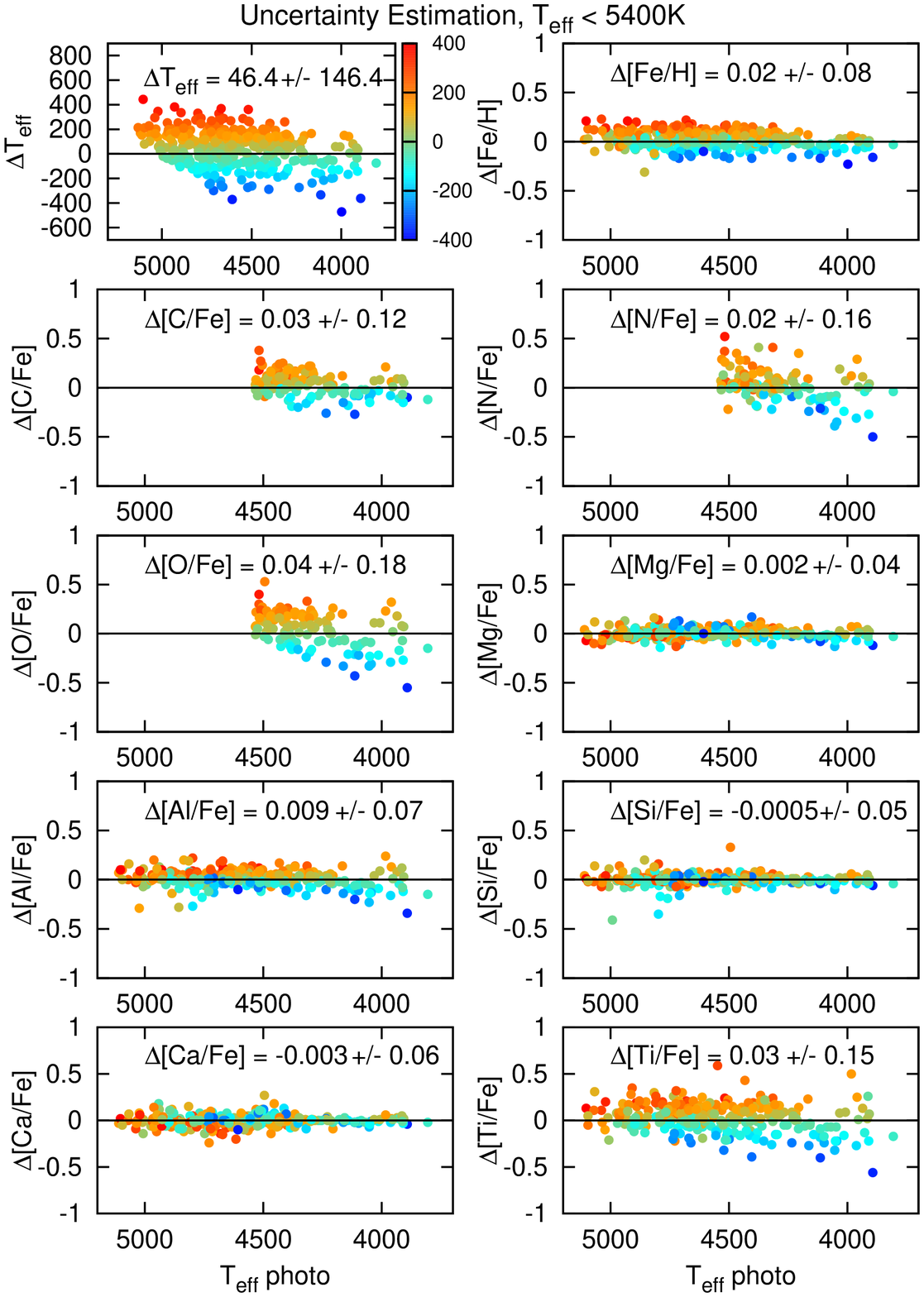}
\caption{Differences in abundances produced by two runs adopting different temperatures: photometric 
and ASPCAP temperatures; otherwise the same calculation method was used. The points are color 
coded by the differences between the photometric and ASPCAP temperatures. The $\pm$ errors give 
the standard deviation around the mean of the differences.
}
\label{fig:errors}
\end{figure}

\begin{deluxetable*}{lrrrrrrrrrrrrll}
\tabletypesize{\scriptsize}
\tablecaption{Identifiers, stellar parameters and elemental abundances from the literature for stars in our sample}
\tablewidth{0pt}
\tablehead{
\colhead{2MASS ID} & 
\colhead{Cluster} & 
\colhead{T$_{\rm eff}$} & 
\colhead{$\log~g$} &  
\colhead{[Fe/H]} & 
\colhead{[C/Fe]} &
\colhead{[N/Fe]} & 
\colhead{[O/Fe]} & 
\colhead{[Mg/Fe]} & 
\colhead{[Al/Fe]} & 
\colhead{[Si/Fe]} &
\colhead{[Ca/Fe]} &
\colhead{[Ti/Fe]} &
\colhead{alt. ID} &
\colhead{Source} }
\startdata
2M21290843+1209118 & M15 & \nodata & \nodata & -2.37 & \nodata & \nodata & \nodata & \nodata & \nodata & 0.36 & \nodata & 0.31 & B5 & h \\
2M21293353+1204552 & M15 & \nodata & \nodata & -2.37 & \nodata & \nodata & \nodata & \nodata & \nodata & \nodata & 0.13 & \nodata & K22 & h \\
2M21293871+1211530 & M15 & \nodata & \nodata & -2.37 & \nodata & \nodata & \nodata & \nodata & \nodata & \nodata & 0.14 & \nodata & K47 & h \\
2M21294979+1211058 & M15 & \nodata & \nodata & -2.40 & \nodata & \nodata & \nodata & \nodata & \nodata & 0.53 & 0.28 & \nodata & K144 & h \\
2M21294979+1211058 & M15 & \nodata & \nodata & -2.27 & \nodata & \nodata & 0.42 & \nodata & \nodata & \nodata & 0.38 & \nodata & K144 & i \\   
\enddata
\tablecomments{This table is available in its entirety in machine-readable form in the online journal. A portion 
is shown here for guidance regarding its form and content. 
Individual star references: (a) \citet{connell01}; (b) \citet{carretta02}; (c) \citet{johnson01};
(d) \citet{sneden01}; (e) \citet{yong01}; (f) \citet{cohen01}; (g) \citet{cavallo01}; (h) \citet{sneden02};
(i) \citet{minniti01}; (j) \citet{otsuki01}; (k) \citet{sneden04}; (l) \citet{sneden05}; (m) \citet{sobeck01};
(n) \citet{kraft01}; (o) \citet{kraft02}; (p) \citet{johnson02}; (q) \citet{lai02}; (r) \citet{ivans01};
(s) \citet{koch01}; (t) \citet{sneden03}; (u) \citet{ram01}; (v) \citet{yong02}; (w) \citet{mel02};
(x) \citet{briley01}; (y) \citet{shetrone01}; (z) \citet{smith02}; (aa) \citet{lee01}; (ab) \citet{ramirez01};
(ac) \citet{yong03}; (ad) \citet{sneden02}; (ae) \citet{roederer01}}
\end{deluxetable*}

The differences in abundances as a function of photometric temperatures are demonstrated in 
Figure \ref{fig:errors}. The top left panel displays the differences in the measured abundances 
by using ASPCAP and photometric temperature, while the rest of the panels are assigned to each element. 
The color scale in all panels represents $\Delta$~ T$_{\rm eff}$. 
We defined the estimated errors associated with the atmospheric parameters based on 
the standard deviation around the mean differences between the two 
temperature scales. The calculated standard deviation in of the differences in temperatures is
146~K (which we round to 150~K). This standard deviation corresponds to the sum of the uncertainty 
in the photometric temperature and the ASPCAP temperature in quadrature. 

In order to estimate the uncertainty of just the photometric temperature component, 
we carried out calculations of temperatures
with varied J$-$Ks colors, reddenings and metallicities for M107. M107 was chosen because it 
is the cluster with the highest reddening in our sample.  The uncertainty of 2MASS photometry is 
usually between 0.025$-$0.03 magnitude for these stars, and by using 0.03 as baseline, 
we estimate an uncertainty of 0.05 magnitude for the J$-$Ks color. 
Changing J$-$Ks by 0.05 magnitudes typically produces a change of 80~K in the photometric temperature.
The reddening of M107 was changed by 0.03, simulating a 10\% uncertainty in reddening, and this produced a difference of about 
40~K in the photometric temperature. A change of 0.1~dex in metallicity results in 1~K or less uncertainty in temperature; thus 
the uncertainties in metallicity can be neglected.  By adding the uncertainty from photometry and reddening in 
quadrature, we estimate the uncertainty of the photometric temperature to be about 100~K.  
The top panel of Table 4 lists uncertainties associated with both 150 and 100~K changes in temperature. 

In our methodology a 100~K change in temperature also introduces an 0.3~dex 
systematic difference in surface gravity, and 0.1~km/s in microturbulent velocity, so by coupling these two 
parameters to T$_{\rm eff}$, our systematic uncertainties include the investigation of abundance 
sensitivity to these parameters as well.  

\subsubsection{Internal Uncertainties}

Besides the uncertainty coming from the adopted atmospheric parameters, the uncertainty of the fit was 
also calculated using the $\sigma$ of the residuals between the best fit synthesis and the observation. These calculations 
estimate random uncertainties. This $\sigma$ of the fit is calculated within the windows used in the $\chi^2$ calculation. 
For determining the uncertainty 
of the fit, we multiply the observed spectra by 1+$\sigma$ and 1-$\sigma$ which simulates two 
spectra slightly different from the original spectrum. Then, the fit of each line is repeated while 
keeping all other parameters unchanged. The average differences between these two new fits and 
the original best-fit spectrum is the defined uncertainty associated with the fit itself. 
This uncertainty estimate is mainly sensitive to variations of noise in the spectrum in the 
defined windows for the $\chi^2$ fit.  If its value is close to 0, then the 
$(1+\sigma) \times$ spectrum and $(1-\sigma) \times$ spectrum give very similar, or the same abundances. 
This is expected when working at high S/N and with a well defined continuum.

\begin{figure*}[!ht]
\centering
\includegraphics[width=5in,angle=0]{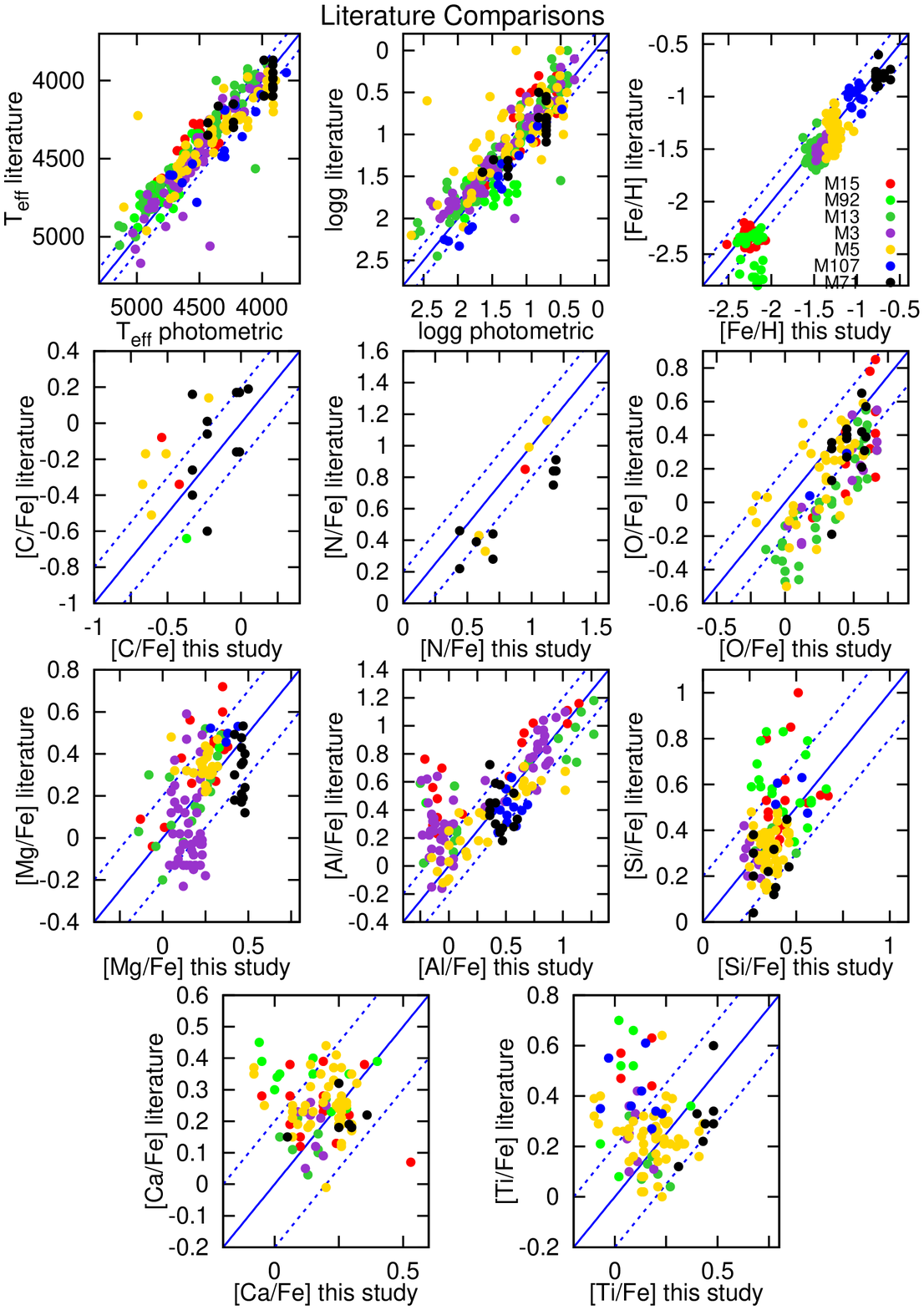}
\caption{T$_{\rm eff}$, $\log~g$, [Fe/H], [C/Fe], [N/Fe], [Mg/Fe], [Al/Fe], [Si/Fe], [Ca/Fe], [Ti/Fe] compared to literature sources. 
Different colors denote different clusters, for an explanation of the colors see upper right panel. Solid lines show the 1:1 
relation, while dashed lines denote $\pm$100~K for T$_{\rm eff}$, and $\pm$0.2~dex for $\log~g$ and 
individual abundances. A detailed discussion can be found in Section 4. 
}
\label{fig:litcomp1}
\end{figure*}

While this uncertainty estimation method is reliable in most cases, it has its limitations for very noisy 
spectra, very weak lines, or when abundances are near upper limits. Uncertainties are usually overestimated for noisy 
spectra, while they are 
underestimated for weak lines and upper limits. Thus, we decided to use one uncertainty from 
\textit{autosynth} per element per cluster by simply averaging together all uncertainties reported by the program. 
This resulted in one uncertainty estimate for every element in each cluster. 
\textit{Autosynth} uncertainties are listed in the middle section of Table 4. The final estimated uncertainties 
were calculated by adding together in quadrature the uncertainties for 100~K difference in temperature (also 0.3~dex 
systematic difference in $\log~g$ , and 0.1~km/s in microturbulent velocity), and the average \textit{autosynth} values 
that estimate random uncertainties. These final estimations for each element per cluster are given in 
the bottom panel of Table 4.

\section{Literature Comparisons}

We take [X/Fe] abundance values for C, N, O, Mg, Al, Si, Ca and Ti from high-resolution spectroscopic 
studies in the literature as a point of comparison for our abundance determinations. 
We use the same literature sources as \citet{meszaros02}, and added more recently 
published papers listed in Table 5. The derivation of stellar parameters 
$\rm{T}_{\rm eff}$ and $\log~g$ are described in detail in Section 2; and they were also 
compared to the literature. Our goal for these 
comparisons is not to cross-calibrate our new abundance determinations with the literature; rather, 
we are looking for cases where our abundances are systematically different from the literature, or 
particular clusters or elements for which our homogeneously observed and analysed data set can 
clarify conflicts in the literature.

Cross-identification between the globular cluster stars in the DR10 APOGEE data set and the literature was 
performed using the Simbad online service\footnote{http://simbad.u-strasbg.fr/simbad/}, based on 2MASS 
identifiers and ($\alpha$, $\delta$) coordinates. Because there is a large and heterogeneous literature 
on chemical abundances in globular cluster stars, we are providing our cross-identifications as a resource 
for the community. Table 5 lists 2MASS ID and position, and alternate stellar identifiers from literature 
abundance studies for all of the stars considered in this study.

\begin{deluxetable*}{lrrrrrrrrrr}
\tabletypesize{\scriptsize}
\tablewidth{0pt}
\tablecaption{Abundance Averages and Scatter}
\tablehead{
\colhead{Cluster} & \colhead{[Fe/H]} & 
\colhead{[C/Fe]} & \colhead{[N/Fe]} &
\colhead{[O/Fe]} & \colhead{A(C+N+O)} & \colhead{[Mg/Fe]} &
\colhead{[Al/Fe]} & \colhead{[Si/Fe]} &
\colhead{[Ca/Fe]} & \colhead{[Ti/Fe]}
}
\startdata
\cutinhead{Averages}
M15 	& -2.28 & -0.41 & 0.95 & 0.54 & 7.09	& 0.11 & 0.34 & 0.44 & 0.16 & 0.15 \\
M92 	& -2.23 & -0.41 & 0.93 & 0.58 & 7.19	& 0.14 & 0.42 & 0.45 & 0.10 & 0.09 \\
M53 	& -1.95 & -0.50 & 1.06 & 0.56 & 7.49	& 0.11 & 0.37 & 0.41 & 0.23 & 0.28 \\
N5466 	& -1.82 & -0.56 & 0.84 & 0.63 & 7.60	& 0.14 & -0.24 & 0.29 & 0.04 & 0.29 \\
M13 	& -1.50 & -0.53 & 0.89 & 0.28 & 7.69	& 0.13 & 0.61 & 0.40 & 0.26 & 0.20 \\
M2 		& -1.49 & -0.48 & 0.90 & 0.41 & 7.76	& 0.26 & 0.45 & 0.35 & 0.24 & 0.27 \\
M3 		& -1.40 & -0.46 & 0.69 & 0.40 & 7.84	& 0.15 & 0.21 & 0.30 & 0.12 & 0.11 \\
M5 		& -1.24 & -0.46 & 0.76 & 0.27 & 7.85	& 0.23 & 0.36 & 0.34 & 0.20 & 0.26 \\
M107 	& -1.01 & -0.21 & 0.69 & 0.33 & 8.15	& 0.24 & 0.47 & 0.48 & 0.15 & 0.21 \\
M71 	& -0.68 & -0.10 & 0.91 & 0.51 & 8.65	& 0.38 & 0.51 & 0.39 & 0.21 & 0.42 \\
\cutinhead{Scatter} 
M15 	& 0.10 & 0.13 & 0.35 & 0.19 & 0.14	& 0.24 & 0.52 & 0.16 & 0.25	& 0.08 \\
M92 	& 0.10 & 0.11 & 0.23 & 0.19 & 0.13	& 0.23 & 0.48 & 0.12 & 0.17 & 0.20 \\
M53 	& 0.07 & 0.16 & 0.21 & 0.06 & 0.09	& 0.08 & 0.51 & 0.05 & 0.17	& 0.13 \\
N5466 	& 0.08 & 0.01 & 0.10 & 0.04 & 0.08	& 0.06 & 0.35 & 0.09 & 0.25	& 0.14 \\
M13 	& 0.07 & 0.07\tablenotemark{a} & 0.17	& 0.17 & 0.15 & 0.15 & 0.53 & 0.09 & 0.14	& 0.14 \\
M2 		& 0.08 & 0.05\tablenotemark{a} & 0.15	& 0.21 & 0.23 & 0.07 & 0.51 & 0.06 & 0.12	& 0.08\tablenotemark{a} \\
M3 		& 0.08 & 0.08\tablenotemark{a} & 0.13	& 0.23 & 0.15 & 0.06 & 0.43 & 0.04 & 0.08	& 0.13 \\
M5 		& 0.08 & 0.13\tablenotemark{a} & 0.17	& 0.27 & 0.19 & 0.09 & 0.34 & 0.07 & 0.11	& 0.12\tablenotemark{a} \\
M107 	& 0.06 & 0.09 & 0.27 & 0.15 & 0.15	& 0.10\tablenotemark{a} & 0.15 & 0.05 & 0.21	& 0.16 \\
M71 	& 0.07 & 0.13 & 0.32 & 0.09 & 0.13	& 0.05\tablenotemark{a} & 0.08 & 0.05 & 0.10	& 0.09 \\
    
\enddata
\tablenotetext{a}{Standard deviation around the linear fit.}
\end{deluxetable*}

Figure \ref{fig:litcomp1} shows comparisons of our stellar parameters and 
abundances against the literature values. Different globular clusters are represented by different colored points. 
In general, we find a systematic offset of $\sim 100$ K between our photometric ${\rm T}_{\rm eff}$ values 
and the spectroscopic effective temperatures from the literature, with a reasonably small scatter. 
This is similar to the typical difference between spectroscopic and photometric temperatures reported by \citet{meszaros02}. 
Because of the degeneracies in deriving stellar parameters, the slightly higher temperatures in the literature 
are accompanied by a systematic offset of $\sim +0.2$~dex in $\log~g$.

There are a few systematic differences between our abundance results and those in the literature. 
These can mainly be traced back to a change in the Solar abundance scale as derived by \citet{asplund01}. 
As one example, the [Fe/H] metallicities we derive are typically $0.1$ dex higher than those from the 
literature, which is quite similar in magnitude to the change in A(Fe)$_{\odot}$ from $7.52$ \citep{anders01} to $7.45$ 
\citep{asplund01}. Also, the Solar abundance of oxygen was revised significantly, 
from A(O)$_{\odot}$=8.93 \citep{anders01} to 8.66 \citep{asplund01}. Since we use the more recent Solar abundance 
values from \citet{asplund01} whereas our earlier literature sources do not, we expect the $\sim 0.3$ dex offset 
visible in Figure \ref{fig:litcomp1}. The systematic difference in [N/Fe] abundance is also likely to be due 
to the updated Solar abundances.

Carbon is the only element studied in this paper for which this explanation does not hold: 
although the Solar abundance was revised down, from A(C)$_{\odot}$=8.56 \citep{anders01} to 
$8.39$ \citep{asplund01}, our [C/Fe] values are typically lower than the literature values. 
Unfortunately, literature values are available for only a subset of our stars, 
which makes it difficult to verify precisely any systematic behavior.

In the other elements we consider in the present study, there is generally good agreement between our 
results and those from the literature, which is encouraging. However, we find significant differences 
in Al abundances in stars with below-Solar metallicities. This is mainly driven by stars in M3, where 
the [Al/Fe] abundances from \citet{johnson02} are larger than in the other literature sources. We 
also see significant scatter in [Ca/Fe] and [Ti/Fe] at low abundances; this is likely caused by our abundance 
determination method having difficulty with weak lines.

The [Mg/Fe] versus [Al/Fe] relations in M3, M13 and M5 provide useful examples of how our new data set 
compares with the literature. In M3, our abundance determinations show a clear bimodality in the 
Mg-Al distribution. The study of \citet{cavallo01} found a similar bimodality, but \citet{johnson02}
found a smooth distribution. In M13 we find an extended anticorrelation between [Mg/Fe] and [Al/Fe], 
and the literature abundances lie within it. We have seven stars in common with \citet{sneden01}, and 
they span the full anticorrelation, while the three and two stars, respectively, that we have in 
common with \citet{cohen01} and \citet{cavallo01} are consistent with our abundance results 
but happen to inhabit small regions of abundance space. M5 is a similar case, where our [Al/Fe] values 
span a range of $-$0.3~dex to $+$1.1~dex, while the \citet{carretta02} Al abundances are limited 
to between $-$0.2 and $+$0.7~dex. As in M13, our sample is larger than 
the literature sample, and covers the full RGB, while the 
literature sample does not fully span the parameter space.

\section{Variations in Individual Element Abundances}

\begin{figure*}[!ht]
\centering
\includegraphics[width=4.5in,angle=270]{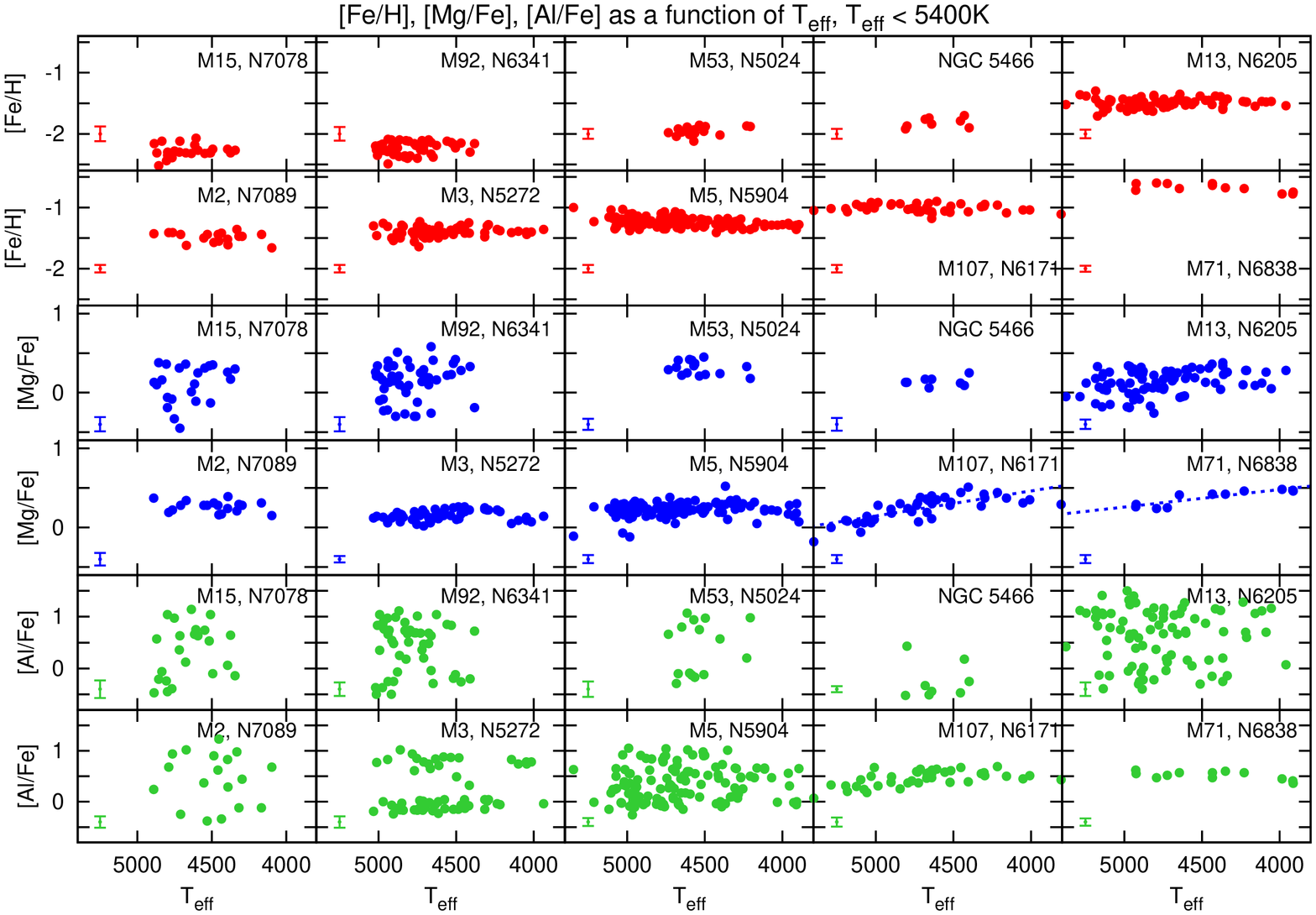}
\caption{[Fe/H], [Mg/Fe], and [Al/Fe] as a function of photometric T$_{\rm eff}$ for all ten clusters. 
The error bars represent our final combined uncertainties from Table~4. The linear fit is 
plotted over [Mg/Fe] for M107 and M71 to remove the visible linear trend (see Section 5.1 for discussion). 
}
\label{fig:femgal}
\end{figure*}

\begin{figure*}[!ht]
\centering
\includegraphics[width=4.5in,angle=270]{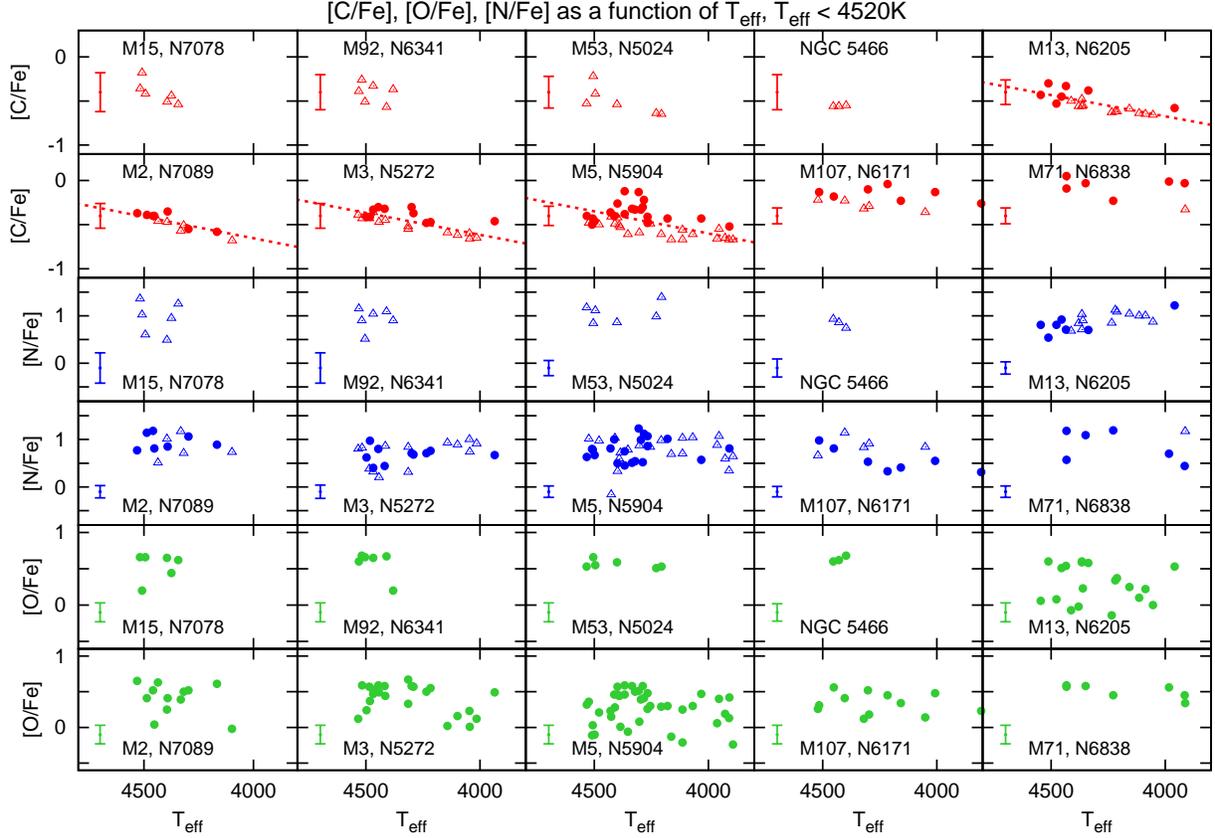}
\caption{[C/Fe], [N/Fe], and [O/Fe] as a function of photometric T$_{\rm eff}$ for all ten clusters. 
Open triangles mark upper limits for [C/Fe] and lower limits for [N/Fe], while the 
real detections are plotted using filled red dots.
The error bars represent our final combined uncertainties from Table~4. The linear correlation in [C/Fe] as a function 
of T$_{\rm eff}$ in M13, M2, M3 and M5 is the effect of CNO burning on the RGB. The fitted lines are used 
to remove the trend in order to estimate the scatter in these clusters (see Section 5.1 for discussion). 
}
\label{fig:cno}
\end{figure*}

\begin{figure*}[!ht]
\centering
\includegraphics[width=4.5in,angle=270]{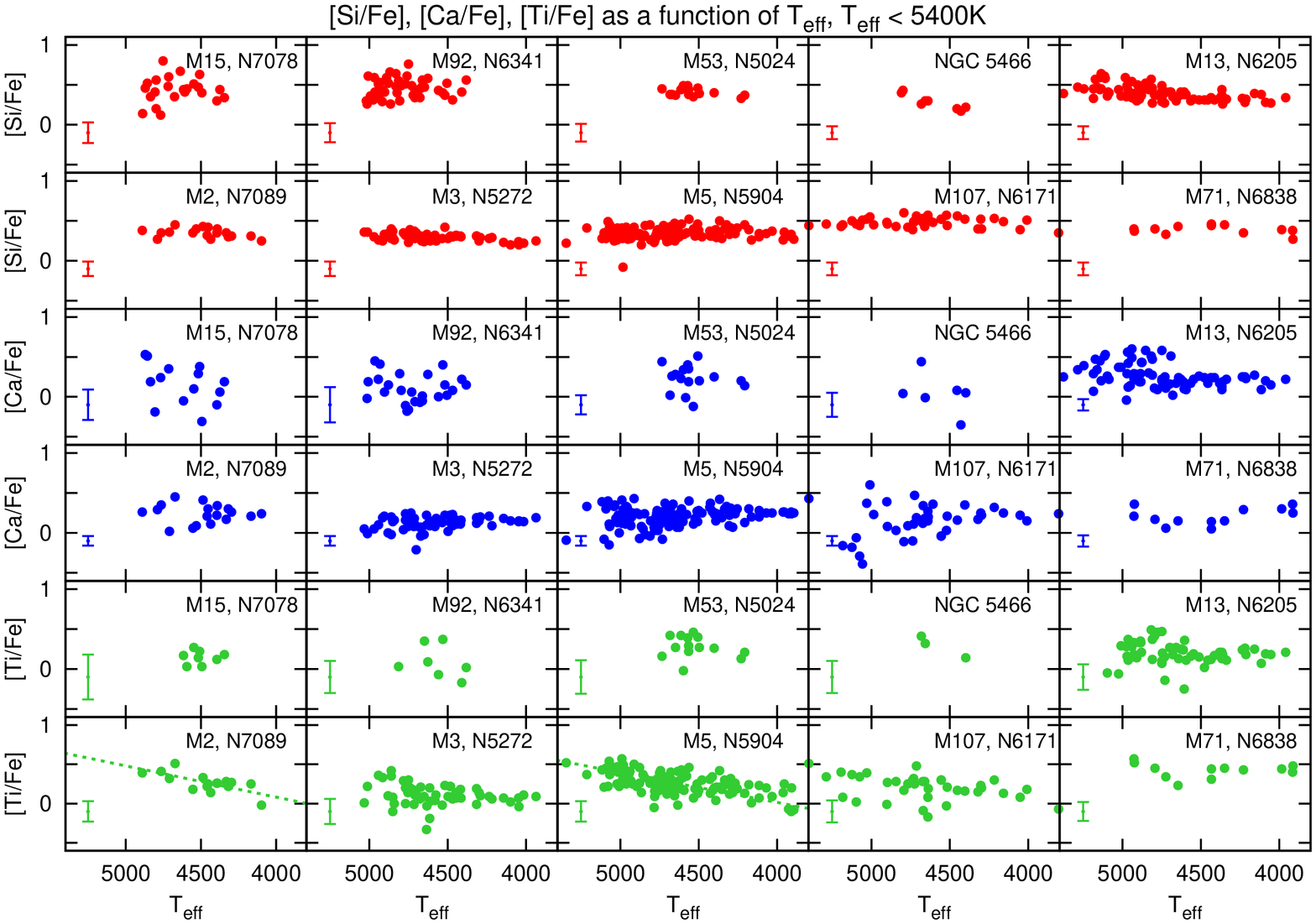}
\caption{[Si/Fe], [Ca/Fe], and [Ti/Fe] as a function of photometric T$_{\rm eff}$ for all ten clusters. 
The error bars represent our final combined uncertainties from Table~4. Trends in Ti for M2 and M5 are removed 
using the plotted lines when calculating the internal scatter (see Section 5.1 for discussion).
}
\label{fig:sicati}
\end{figure*}

Although there are well-known abundance patterns within globular clusters, 
large homogeneous studies that include a wide range of abundances are rare. 
Since we can determine abundances of most of the light elements for the stars 
in our sample, we examine the behavior of the known abundance patterns across 
a wide range in cluster metallicity, and search for unexpected variations in 
the $\alpha-$elements. In this section we focus on the range of abundance 
within each cluster, to separate real abundance variations, bimodalities 
and trends from possible measurement errors. Table 6 lists the average and 
standard deviation for each elemental abundance in each cluster. 

\subsection{Correlations with T$_{\rm eff}$}

In Figures \ref{fig:femgal}-\ref{fig:sicati} we show all nine derived abundances as a function of effective temperature. 
From Figure \ref{fig:femgal} we conclude that we measure constant Fe abundances in all of the clusters. 
Mg and Al abundances show a large range of values in some clusters (as discussed in Section 2, and further in 
Section 6), but no significant trends with T$_{\rm eff}$ except in M107 and M71. This trend is very weak in M71, 
and data are consistent with showing no trend within the uncertainties. However, in M107 the trend is stronger. We 
currently do not fully understand where these small correlations come from. We 
suspect that this is a result of a combination of effects. One such effect may be the use of 
model atmospheres that assume local thermodynamic equilibrium (LTE), but that non-LTE effects 
\citep{bergemann01} act to make the strong Mg lines in these 
metal-rich stars give the appearance of higher abundance in the cooler stars. Other possible effects are small 
systematic errors from estimating T$_{\rm eff}$ and $\log~g$ culminating during the synthesis. Abundances are 
also sensitive to the microturbulent velosity, so if the 
$\log~g-$v$_{micro}$ expression used is not so accurate in this metallicity range, or 
the surface gravity is badly estimated, this propagates into systematically off abundances through v$_{micro}$.

The minimum [C/Fe] (Figure \ref{fig:cno}) that can be measured from the CO lines strongly depends on temperature. 
Because RGB stars in GCs generally have low carbon abundances, we can only set upper limits for a number of our 
stars. Our [C/Fe] values are an average of the derived abundances 
from five CO windows. As a result determining the upper limit is more challenging because CO lines in certain 
windows disappear faster with rising temperature than in others. We carefully checked every CO window fit and selected upper limits 
if the CO band head was not visible in more than three windows by looking at the flatness of 
the $\chi^2$ fit around the minimum value. Because the derived abundance of N from the CN lines is 
anticorrelated with the value of [C/Fe] used, all stars with upper limits in [C/Fe] have also lower 
limits in [N/Fe]. These upper limits for [C/Fe] and lower limits of [N/Fe] are identified with open 
triangles in all figures. 

We see clear correlations between [C/Fe] and T$_{\rm eff}$ in M13, M3, M2, and M5 when omitting 
upper limits. 
We interpret these as a sign of ``deep mixing'', a nonconvective mixing process that causes 
steady depletion of surface carbon abundance and an enhancement in nitrogen abundance in all 
low-mass RGB stars \citep{gra04}. We derive only upper limits on [C/Fe] for M15, M92, M53 and NGC 5466 
because the CO bands are quite weak at such low metallicity. With only upper limits, it is 
not possible to identify abundance trends in these clusters. However, previous studies 
\citep{angelou01, shetrone02, martell01} have found signs of 
the same deep mixing process in action in those clusters. 

The only cluster in which we see a sign of nitrogen enrichment with declining 
temperature is M13. Possible trends in nitrogen are not necessarily a result of astrophysics, 
because in our methodology small systematic errors in temperature can result in systematic errors in N 
abundance because deriving [N/Fe] is challenging from the CN lines. Other than the relatively large errors of N, 
one also has to be careful with possible correlations of C and N with temperature, because they can be  
generated on the RGB as part of deep mixing and also in a previous stellar evolution event 
that are responsible for the pollution.

In Figure \ref{fig:sicati} there is a weak trend, on par with the error, visible in the Si 
abundance in M13, but not in the other clusters. 
While Ca does not show any correlation with temperature, it does show temperature-dependent
scatter in most of the GCs. The three Ca lines used 
in our analysis are generally weak, and get significantly weaker above 4700~K, which leads  
to higher errors related to increased sensitivity to the uncertainty in the continuum placement.
Titanium, unexpectedly, shows a decline with decreasing temperature in M2, M5 and (marginally) in M107 in Figure \ref{fig:sicati}. 
We suspect that inaccuracies in our analysis of the hotter star spectra are driving this apparent trend, since the 
S/N is lower for those stars than for the cooler, brighter giants. Because these trends in Mg (Figure \ref{fig:femgal}), 
Si and Ti (Figure \ref{fig:sicati}) only show up in a handful of clusters, we believe that they are a result 
of difficulties in data analysis for certain lines in certain stars and not any systematic mishandling in, 
e.g., ${\rm v}_{\rm micro}$ estimation.

We choose to fit these various observed trends with a linear equation, regardless of their origin, in order to explore 
the scatter around the trend. We fit lines to [Mg/Fe] 
in M107 and M71, [C/Fe] in M13, M2, M3 and M5, and [Ti/Fe] in M2 and M5. 
We are only using these fits to determine the internal 
scatter, which we define as the standard deviation about the fitted line. 
Throughout the rest of this paper we use the abundance values directly, except when discussing the CN anticorrelations in Section 6.4. 

\subsection{Scatter and Errors}

\begin{figure*}[!ht]
\centering
\includegraphics[width=4.5in,angle=270]{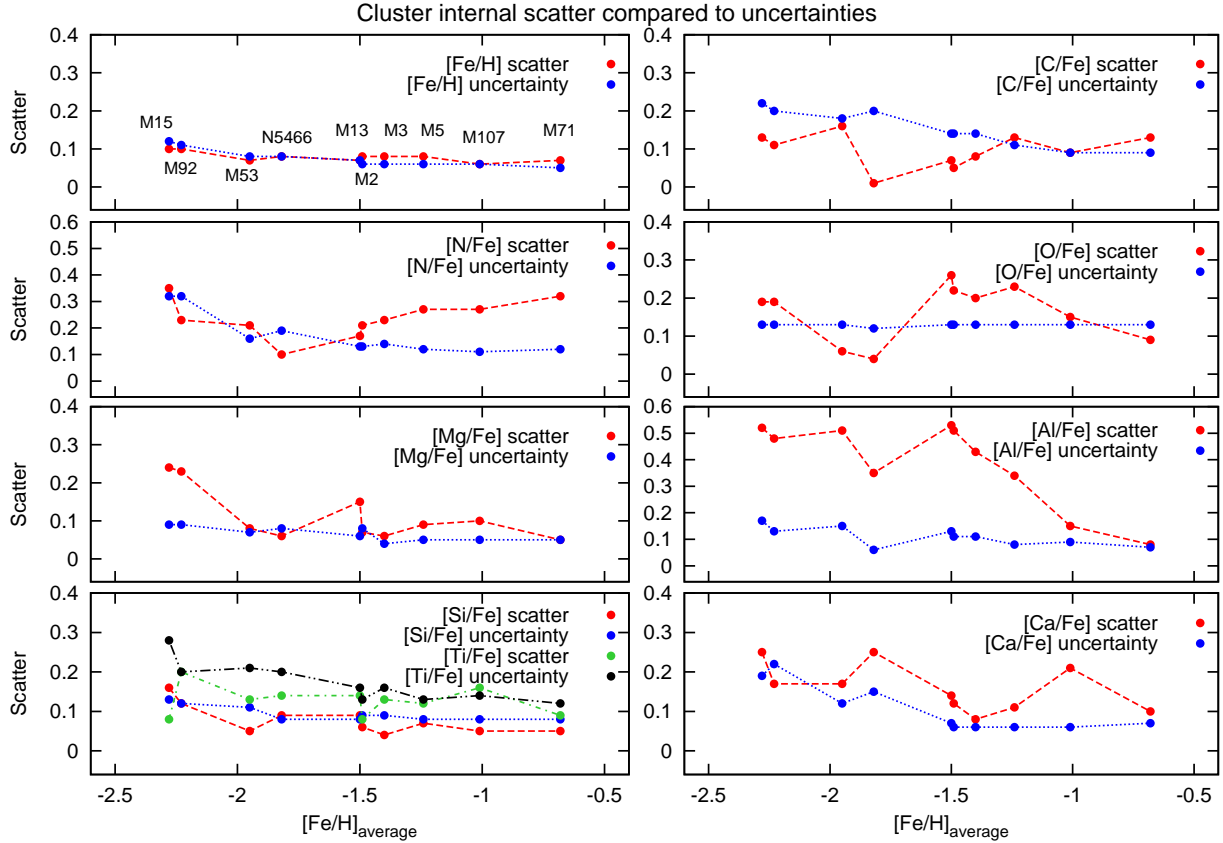}
\caption{The internal scatter in each cluster from Table 6 (red dots) compared to our estimated final 
combined errors from Table 4 (blue dots). 
}
\label{fig:rms}
\end{figure*}

An internal abundance scatter significantly larger than the estimated errors suggests an astrophysical origin. 
In fact, significant scatter in N, O, Mg and Al is well documented in literature \citep{gratton02}. The large star-to-star 
variations in these elements are the result of the CNO, Ne-Na, and Mg-Al cycles, which play an important 
role in the nuclear fusion processes at the early stages of RGB (CNO) and AGB (Ne-Na, Mg-Al) evolution. 
Figure \ref{fig:rms} shows the internal scatter calculated as the standard deviation around the mean values 
(in some cases around a fitted linear equation, as explained earlier), and the final combined estimated uncertainties from Table 4, 
as a function of cluster average metallicity. We note that the small number of stars analysed 
in NGC~5466 limits our ability to make a detailed analysis of C, N, and O in this cluster: because we were only able to 
measure the abundance of these elements in the three stars that have temperatures below $4500$K, the scatter is 
unrealistically higher than the calculated uncertainty. 

The star-to-star scatter in [Fe/H] is quite similar to our measurement uncertainties, indicating that no significant 
metallicity variations in the clusters are detected in this study. Recently, \citet{yong06} have discovered three 
distinctive groups with different iron abundances in M2. The first group has a dominant peak at [Fe/H]=-1.7, while 
the second and third have smaller peaks at −1.5 and −1.0, but the membership for the latter group is not conclusive. 
We see no such behaviour in M2. The discrepancy between these two studies is most likely the result of a selection effect. 
\citet{yong06} selected stars that belong to a second RGB \citep{lardo01} within the cluster, while our sample selection 
were based upon previous observations \citep{zasowski01} and all of our stars belong to the main RGB of this cluster.

The $\alpha-$elements Si and Ti also show constant distribution, but the scatter in [Ca/Fe] is larger than our measurement 
uncertainties for most clusters in our sample. We attribute this to an increasing 
inaccuracy in [Ca/Fe] with rising temperature, as discussed in Section 5.1, and not a real range in calcium abundance. 
Little to no range in iron and $\alpha-$elements is what we expect for normal globular clusters, 
although the clusters in which these abundances do vary are an intriguing puzzle \citep[see, e.g., ][]{marino02}. 
M107 stands out as having the largest discrepancy between the scatter of [Ca/Fe] and its estimated uncertainty. However, 
the large internal scatter is not related to astrophysics, but to spectral contamination from telluric OH lines. Due to the cluster's 
radial velocity, two of the three Ca lines lie on top of atmospheric OH lines. Because the telluric correction 
is not perfect, the Ca lines are compromised in a way that is not included in our uncertainty estimation. 

We do see scatter above the level of the uncertainties in some of the light-element abundances. In the case of carbon, 
we find that the scatter around the overall carbon depletion trend for M13, M2, M3 and M5 (as shown by dashed lines 
in Figure \ref{fig:cno}) is fairly small, although our measurement uncertainties get quite large at low metallicity. 
We do not see clear carbon depletion in M107 or M71; however, as discussed in \citet{martell01}, the rate of carbon 
depletion due to deep mixing is lower in higher-metallicity stars.

The range in [N/Fe] abundance (Figure \ref{fig:rms}, second panel from the top, in the left column) in 
M71, M107, M5, M3, M13, and M2 is 
clearly larger than the estimated uncertainties. However, the difference between the scatter and the uncertainty 
decreases with decreasing cluster metallicity, and below [Fe/H]=$-$1.7 (NGC~5466, M53, M92, and M15) 
our derived [N/Fe] values are all lower limits. The decreasing discrepancy is a result of larger uncertainties 
in more metal poor clusters, and the fact that the minimum N abundance of a cluster is slightly increasing as 
the average metallicity decreases (Figure \ref{fig:cno}), 
while the maximum N is constant above [M/H]=-1.7. The uncertainties in M15 and M92 are significantly larger 
than in other clusters, and our measurements are not precise enough to judge the amount of N enrichment 
in these clusters. Fitting the CN lines in 
our defined CN window is also challenging. This is because they are usually blended with other 
molecular lines, and because N abundance strongly depends on the 
derived C abundance, which is an upper limit in many cases. A more detailed analysis of the 
C-N anticorrelation and N bimodality can be found in Section 6.4.

\begin{deluxetable*}{lcrrrrrrrrrrr}
\tabletypesize{\scriptsize}
\tablewidth{0pt}
\tablecaption{Averages of Populations}
\tablehead{
\colhead{Cluster} & \colhead{Pop.\tablenotemark{a}} & 
\colhead{N\tablenotemark{b}} & \colhead{R\tablenotemark{c}} & \colhead{[Fe/H]} &
\colhead{[C/Fe]} & \colhead{[N/Fe]} &
\colhead{[O/Fe]} & \colhead{[Mg/Fe]} &
\colhead{[Al/Fe]} & \colhead{[Si/Fe]} &
\colhead{[Ca/Fe]} & \colhead{[Ti/Fe]}
}
\startdata
M15 	& 1 & 10 & 0.43 & -2.31 & -0.49 & 0.78 & 0.64 & 0.20 & -0.19 & 0.31 & 0.08 & 0.11 \\
		& 2 & 13 & 0.57 & -2.26 & -0.33 & 1.11 & 0.43 & 0.04 & 0.75 & 0.53 & 0.24 & 0.17 \\
M92 	& 1 & 14 & 0.30 & -2.23 & -0.42 & 0.89 & 0.67 & 0.33 & -0.23 & 0.41 & 0.08 & 0.09 \\
		& 2 & 33 & 0.70 & -2.23 & -0.38 & 1.03 & 0.40 & 0.05 & 0.70 & 0.47 & 0.11 & 0.09 \\
M53 	& 1 & 8  & 0.50 & -1.96 & -0.43 & 0.91 & 0.59 & 0.38 & -0.11 & 0.42 & 0.23 & 0.24 \\
		& 2 & 8  & 0.50 & -1.95 & -0.54 & 1.13 & 0.55 & 0.25 & 0.84 & 0.41 & 0.23 & 0.31 \\
N5466 	& 1 & 6  & \nodata & -1.83 & -0.56 & 0.84 & 0.64 & 0.15 & -0.42 & 0.28 & 0.14 & 0.29 \\
		& 2 & 2  & \nodata & -1.79 & -0.56 & 0.86 & 0.62 & 0.11 & 0.31 & 0.30 & -0.16 & \nodata \\
M13 	& 1 & 32 & 0.40 & -1.50 & -0.44 & 0.83 & 0.56 & 0.21 & 0.04 & 0.38 & 0.24 & 0.18 \\
		& 2 & 49 & 0.60 & -1.49 & -0.58 & 0.92 & 0.12 & 0.09 & 0.99 & 0.40 & 0.28 & 0.21 \\
M2 		& 1 & 7  & 0.39 & -1.46 & -0.45 & 0.75 & 0.56 & 0.30 & -0.10 & 0.35 & 0.19 & 0.26 \\
		& 2 & 11 & 0.61 & -1.50 & -0.49 & 1.01 & 0.30 & 0.24 & 0.79 & 0.35 & 0.27 & 0.27 \\
M3 		& 1 & 39 & 0.66 & -1.40 & -0.42 & 0.59 & 0.53 & 0.18 & -0.08 & 0.30 & 0.13 & 0.11 \\
		& 2 & 20 & 0.34 & -1.38 & -0.54 & 0.86 & 0.16 & 0.10 & 0.79 & 0.29 & 0.10 & 0.12 \\
M5 		& 1 & 60 & 0.49 & -1.24 & -0.41 & 0.62 & 0.41 & 0.24 & 0.06 & 0.34 & 0.20 & 0.24 \\
		& 2 & 62 & 0.51 & -1.24 & -0.50 & 0.87 & 0.16 & 0.21 & 0.64 & 0.35 & 0.20 & 0.28 \\
M107\tablenotemark{d} 	& 1 & 6  & \nodata & -1.04 & -0.16 & 0.47 & 0.38 & 0.36 & 0.51 & 0.50 & 0.21 & 0.12 \\
		& 2 & 6  & \nodata & -1.00 & -0.25 & 0.92 & 0.29 & 0.37 & 0.53 & 0.45 & 0.22 & 0.16 \\
M71\tablenotemark{d} 	& 1 & 3  & \nodata & -0.72 & 0.00 & 0.57 & 0.53 & 0.46 & 0.44 & 0.41 & 0.24 & 0.38 \\
		& 2 & 4  & \nodata & -0.69 & -0.17 & 1.16 & 0.49 & 0.44 & 0.53 & 0.38 & 0.21 & 0.45 \\
\enddata
\tablenotetext{a}{Population 1 is denoted by red, while population 2 is denoted by blue in Figures 8$-$12.}
\tablenotetext{b}{N: the number of stars in each population.}
\tablenotetext{c}{R: the ratio of number of stars in a population and the overall number of stars analysed.}
\tablenotetext{d}{Populations are found using N instead of Al.}
\end{deluxetable*}

Our uncertainties of the [O/Fe] abundance (Figure \ref{fig:rms}) 
are fairly constant as a function of cluster 
metallicity. This is because the error is dominated by a strong sensitivity to the temperature used in the model 
atmosphere, and the internal scatter varies between clusters. 
Unfortunately, the Na lines available in the APOGEE spectra are too weak to investigate the O-Na anticorrelation in our sample, so 
we can only limit our discussion on the scatter of [O/Fe]. We use the later as an indication of the strength of O-Na anticorrelation 
in a cluster. In M71 and M107 the O scatter is comparable to the uncertainties, suggesting that the O-Na anticorrelation 
is not as extended as in the rest of the clusters. A clear spread is visible in M5, M3, M2, M13, M15 and M92, while in M53 the 
O abundances are constant and the scatter is significantly smaller than what is expected from the uncertainties.

The effects of the Mg-Al cycle are very noticeable for all clusters except M71 and M107 
(Figure \ref{fig:rms}, second panels from the bottom). The scatter in Mg abundance in those clusters is close to
the estimated uncertainties after taking the linear correlations with 
T$_{\rm eff}$ into account (Figure \ref{fig:femgal}). In other clusters the scatter of [Mg/Fe] 
also closely follows the uncertainties, except for M15 and M92 where the most Mg poor stars can be 
found. We see no Al enrichment in M107 and M71; however, as the average metallicity 
decreases, the amount of internal scatter rapidly starts to deviate from the error, and we see a large 
spread in Al in all other GCs. A more detailed analysis of the Mg-Al anticorrelations and Al populations can be found in 
Section 6.2 and 6.3.

Based on the examination of the spread of N and Al abundances, we conclude that all clusters possess multiple stellar 
populations either based on N or Al, or both. We examine these populations in more detail in the next section. 
The derived $\alpha-$element abundances are also fairly constant in all globular clusters 
presented here, similar to what is reported in the literature \citep{gra04}.

\section{Multiple Populations}

Most light elements show star-to-star variations 
in all GCs. These large variations are generally interpreted as the result of chemical feedback from an earlier 
generation of stars \citep{gratton01, cohen02}, 
rather than inhomogeneities in the original stellar cloud these stars formed from. 
Thus, the current scenario of GC evolution generally assumes that more than one population of stars were formed in each cluster. 
The first generation of GC stars formed from gas that had been enriched by supernovae in 
the very early Universe, while the second formed by combining gas from the original star-forming 
cloud with ejecta from the first-generation stars. Only the fraction of first-generation stars 
contribute to the pollution, and the time scale of the formation of these second 
generation stars depend on the nature of polluters; it is a couple of hundred Myr in the case of 
intermediate-mass AGB stars, but it is only a few Myr for fast rotating massive stars and massive binaries.

These first-generation pollutors are thought to have been fast rotating massive stars \citep{decressin01}, or 
intermediate mass (M$_{\rm star} >$~3~M$_{\odot}$) AGB stars \citep{dercole01}, or massive binaries \citep{demink01}. Our data reveal the expected 
signatures of pollution from material enriched from the hot hydrogen 
burning cycles such as the CNO, Ne–Na, or Mg–Al cycles in all globular clusters in our sample. In this section 
we explore the various correlations between these elements, and we discuss individual star formation, and/or 
pollution events suggested by separate groups found in the C-N and Mg-Al anticorrelations.

\subsection{Identifying Multiple Populations}

To separate the various populations 
under study, we follow an approach similar to that of
\citet{Gratton11a}, who used $K$-means clustering \citep{Steinhaus56}
to identify multiple populations in $\omega$ Cen. Here, we use the
\emph{extreme-deconvolution} (XD) method of
\citet{Bovy11a}\footnote{Code available at \texttt{http://github.com/jobovy/extreme-deconvolution}~.} to
identify population groups and assign membership. This method fits the
distribution of a vector quantity, here the elemental abundances, as a
sum of $K$ Gaussian populations, whose amplitudes, centers and
covariance matrices are left entirely free. The algorithm can be
applied to noisy or incomplete data, thus making the best possible use
of all available data. In the current application, XD's main advantage
is the latter, since we are not able to measure abundances for all nine 
elements in all stars in our sample. Similar to the $K$-means, the 
number $K$ of populations to fit is
an input to the algorithm; XD itself does not determine the optimal
number of components to fit the distribution.

Briefly, the XD algorithm works by optimizing the likelihood of the
Gaussian mixture model of the data. The optimization proceeds by an
iterative procedure consisting of repeated \emph{expectation} (E) and
\emph{maximization} (M) steps. In the E step, each datum is
probabilistically assigned (a) membership in each population and (b)
error-free values of each noisy or missing elemental abundance; in the
M step, each Gaussian's parameters are updated using its members' mean
abundances and covariance calculated from the error-free values
obtained in the E step. These steps are repeated until the likelihood
stops increasing to within a small tolerance. The algorithm is proven
to increase the likelihood in each EM-step.

We analyse each globular cluster in our sample using this algorithm,
with the [Mg/Fe],[Al/Fe], [Si/Fe], [Ca/Fe], [Ti/Fe] abundances as well as only 
[Mg/Fe] and [Al/Fe]. We find that the two separate analyses 
provide nearly identical results, indicating that Si, Ca, and Ti do 
not drive the populations. For M71 and M107 we used [N/Fe] instead of [Al/Fe], 
because Al does not show any spread in these two clusters. We run XD using uncertainties 
given by the $`$Final combined errors$'$ section in Table 4, and also without any 
uncertainties, and found that the two different runs produce identical 
results. We include missing abundance 
measurements by using a large uncertainty for these measurements. Group membership for each star is
determined using the best-fit Gaussian mixture by calculating the
posterior probability for each star to be a member of each population
based on its elemental abundances and their uncertainties; stars are
then assigned to the population for which this probability is the
largest. These membership assignments are typically unambiguous
(probabilities $\gtrsim 99~\%$ in most cases), with only a few cases
for which the maximum probability is below 90~\%. The abundance averages of populations are 
listed in Table 7. In the next few subsections we discuss the results from the population 
fitting in more detail for each individual cluster.

\begin{deluxetable}{lrrrrrr}
\tabletypesize{\scriptsize}
\tablewidth{0pt}
\tablecaption{Statistics of Correlations}
\tablehead{
\colhead{Cluster} & 
\colhead{$\Delta$}\tablenotemark{a} & 
\colhead{e$\Delta$}\tablenotemark{b} & 
\colhead{$\sigma$}\tablenotemark{c} & 
\colhead{$\Delta$} & 
\colhead{e$\Delta$} & 
\colhead{$\sigma$} \\
\colhead{} & 
\colhead{[Mg/Fe]} & 
\colhead{[Mg/Fe]} & 
\colhead{} & 
\colhead{[Si/Fe]} & 
\colhead{[Si/Fe]} & 
\colhead{} 
}
\startdata
M15 	& 0.16 		& 0.037 	& 4.3 		& 0.22	& 0.055	& 4.0 	\\
M92 	& 0.28 		& 0.029 	& 9.7 		& 0.06 	& 0.038	& 1.6 	\\
M53 	& 0.13 		& 0.035  	& 3.7 		& 0.01 	& 0.055	& $<$1	\\
M13 	& 0.12 		& 0.014 	& 8.6 		& 0.02 	& 0.018	& 1.1	\\
M2 		& 0.06 		& 0.038  	& 1.6 		& 0.00	& 0.043	& $<$1	 \\
M3 		& 0.08 		& 0.011 	& 7.3 	 	& 0.01	& 0.024	& $<$1	\\
M5 		& 0.03 		& 0.008 	& 3.8 		& 0.01	& 0.014	& $<$1	\\
\enddata
\tablenotetext{a}{Difference of the average Mg and Si abundance of populations.}
\tablenotetext{b}{The estimated error of the difference.}
\tablenotetext{c}{Detection significance in $\sigma$.}
\end{deluxetable}

\subsection{The Mg-Al anticorrelation}

\begin{figure*}[!ht]
\centering
\includegraphics[width=4.5in,angle=270]{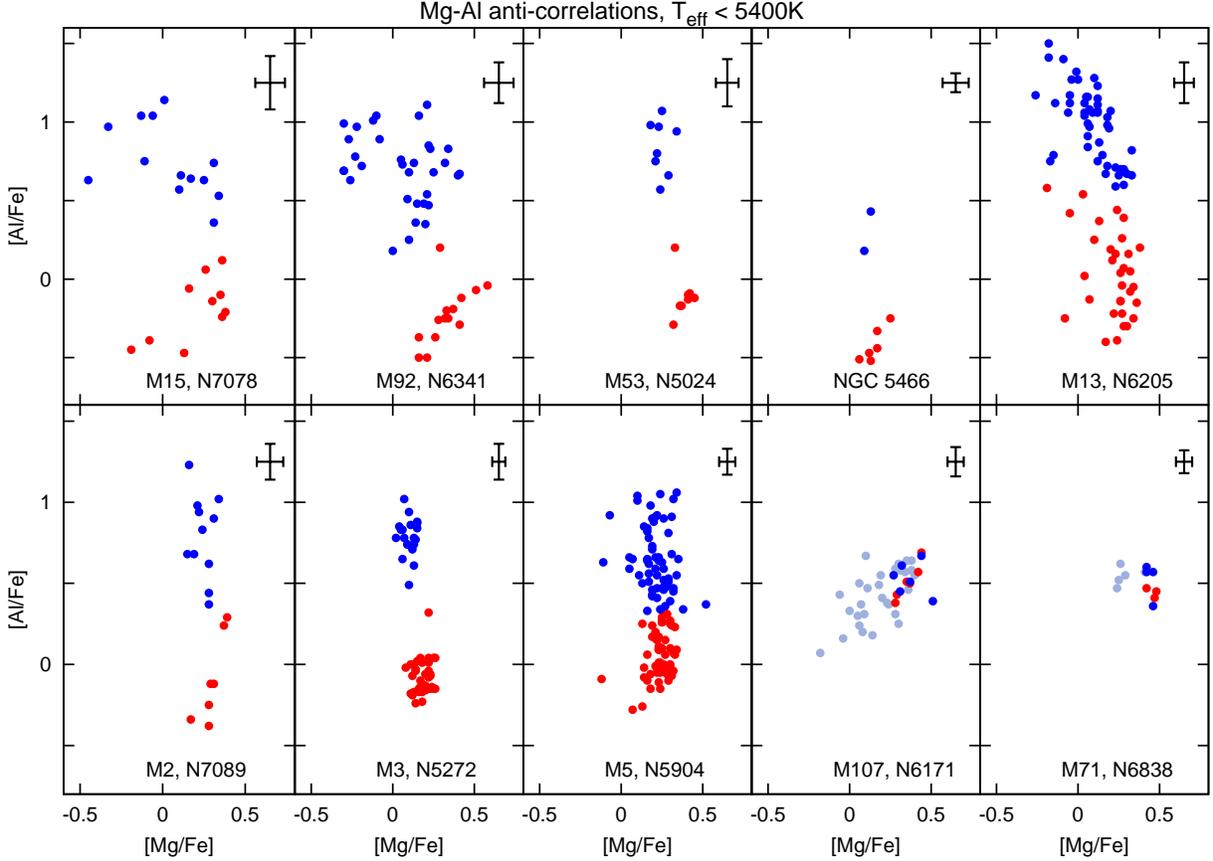}
\caption{Mg-Al anticorrelations. Colors mark 
the populations found with the XD code, first population is denoted by red, and the second-generation denoted by blue points. 
Because [N/Fe] was used to identify populations in M107 and M71, not all stars could be catalogued, these are denoted by light blue.  
The same colors denote the same stellar groupings in 
Figures \ref{fig:almganti}-\ref{fig:cnoadd}.
}
\label{fig:almganti}
\end{figure*}

\begin{figure}[!ht]
\centering
\includegraphics[width=2.4in,angle=270]{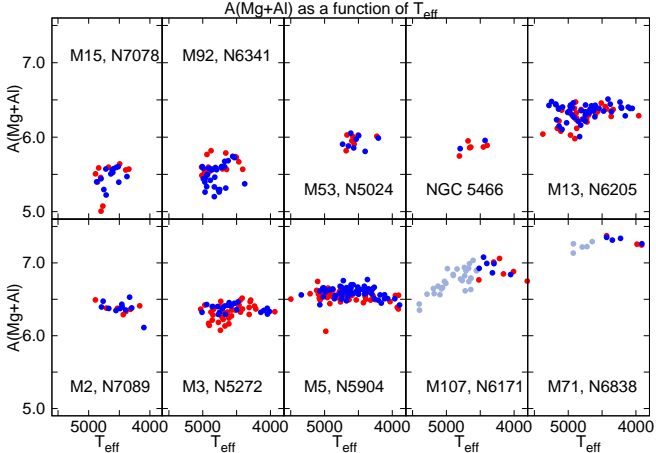}
\caption{The combined abundance of A(Mg+Al) as a function of effective temperature.
}
\label{fig:mgaladd}
\end{figure}

\begin{figure*}[!ht]
\centering
\includegraphics[width=6in,angle=0]{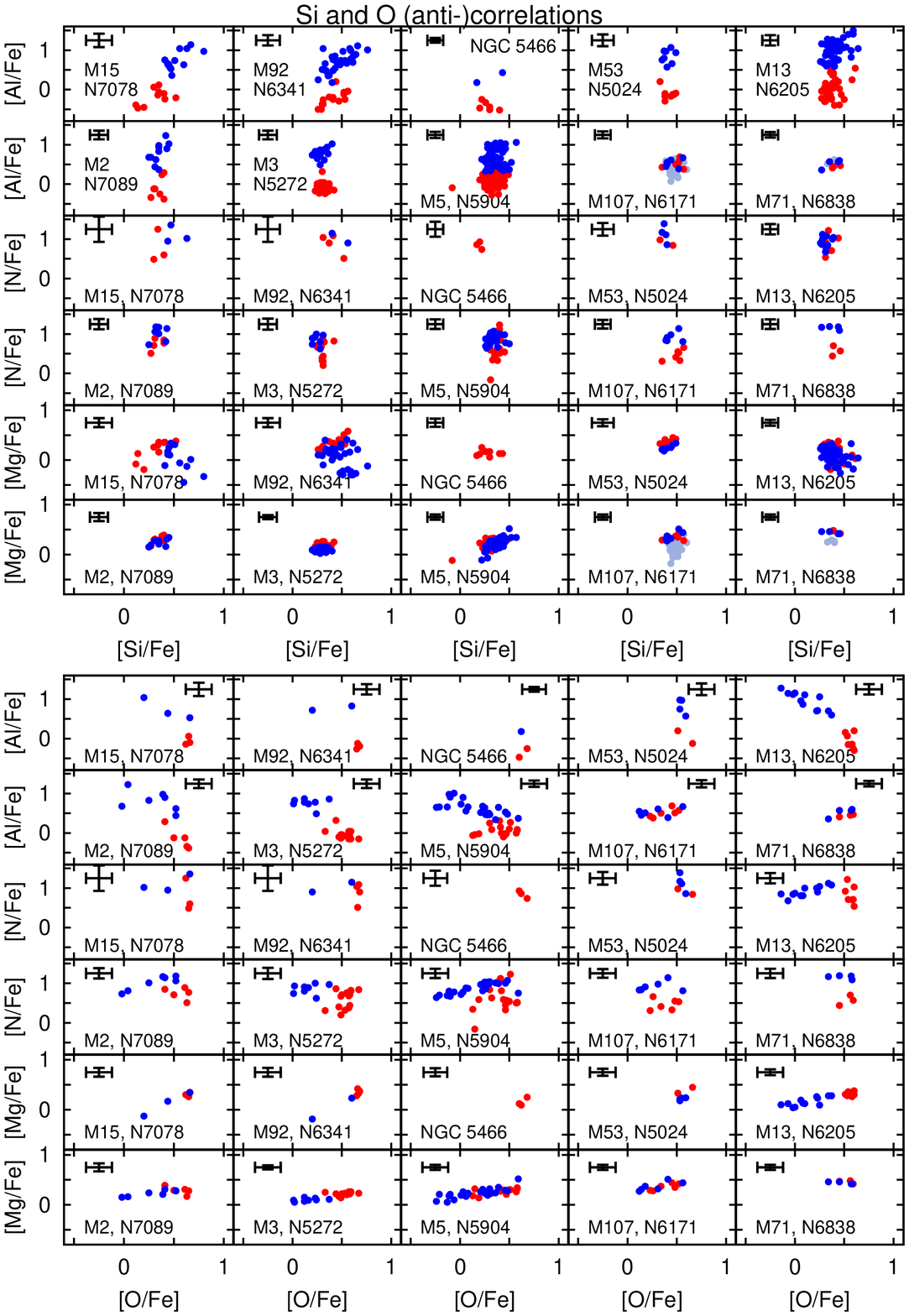}
\caption{Si and O (anti)correlations with Al, N, and Mg. 
For explanation of color coding, please see description of Figure \ref{fig:almganti}.
}
\label{fig:sicor}
\end{figure*}

In order for the Mg-Al cycle to operate, high temperatures above 70 million Kelvin are required \citep{charbonnel01}. 
Because current cluster main-sequence stars are unable to reach these temperatures, the high [Al/Fe] abundances we see in some 
globular cluster stars imply that a previous generation of higher-mass or evolved stars 
must have contributed to their chemical composition.

\begin{figure*}[!ht]
\centering
\includegraphics[width=4.5in,angle=270]{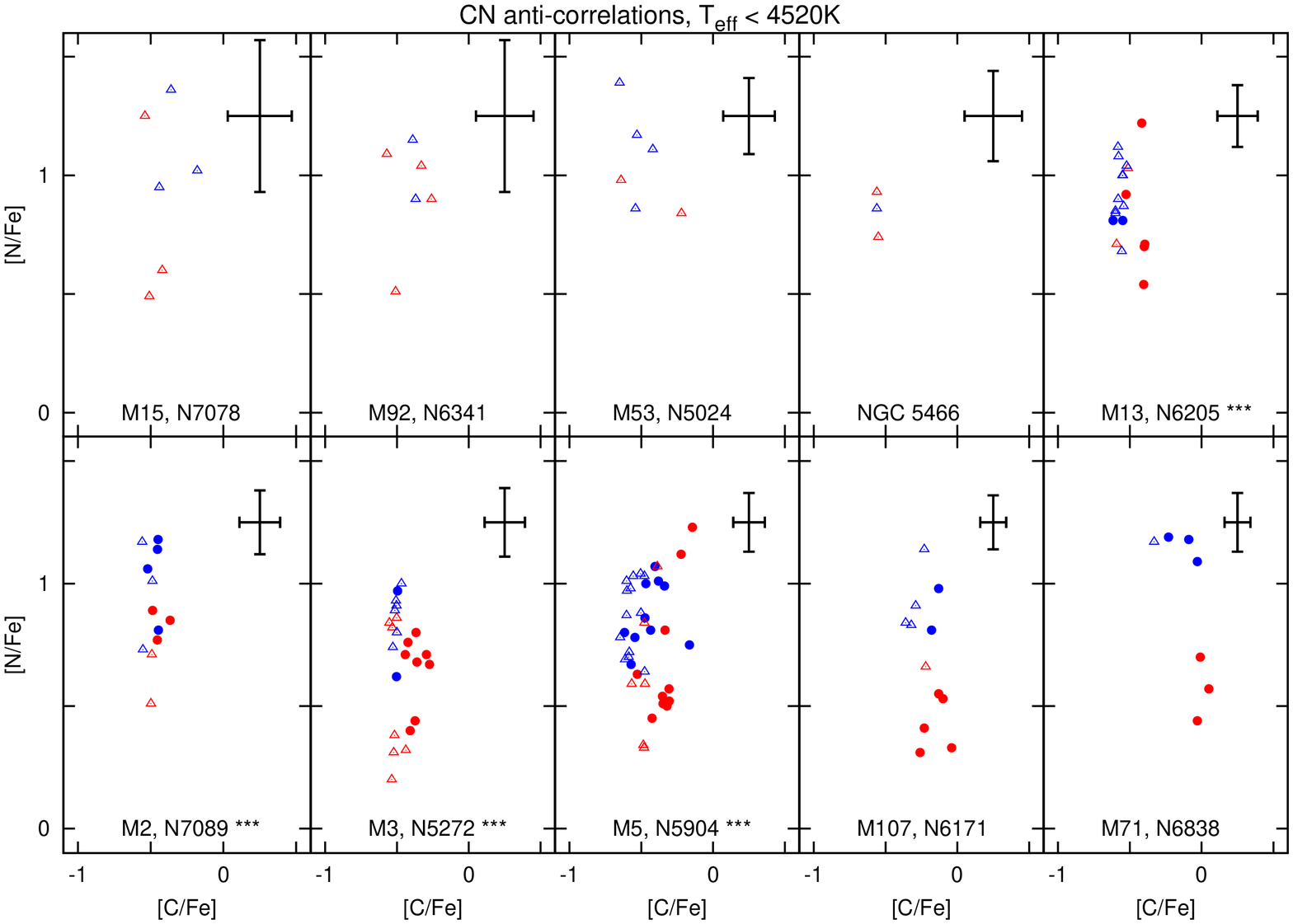}
\caption{CN anticorrelations. Correlations of [C/Fe] with temperature associated with deep mixing were removed 
in clusters marked by ***. Upper limits are denoted by open triangles. 
}
\label{fig:cnanti}
\end{figure*}

Figure \ref{fig:almganti} shows [Mg/Fe] versus [Al/Fe] for all ten clusters in our study. 
The two XD populations are plotted in red and blue, and a representative error bar is 
included in the top right corner of each panel. As noted previously, stars in M107 and 
M71 do not show a strong Mg-Al anticorrelation, and their XD population assignments are 
based on N abundances. The light blue points in the panels for M107 and M71 are 
stars that are warmer than 4500~K, and thus have no CNO measurements.

The extended distribution of Mg and Al abundances apparent in Figure \ref{fig:almganti} 
is typical for globular clusters, and it shows the influence of the Mg-Al fusion cycle, 
which converts Mg into Al. However, there is a variety in the structure of this relationship 
that has not been thoroughly explored before: in some clusters (M92, M53, NGC 5466, M2 and M3), 
there are two distinct abundance groups with a gap, while in other clusters (M15, M13 and M5) 
there is no gap. This result strongly suggests that there is diversity in the process of stellar 
chemical feedback and star formation in globular clusters, which may relate to the larger 
environment in which they formed. 

To date, the largest homogeneous study of Mg and Al abundances was carried out by \citet{carretta02}. 
We have several clusters in common, M15, M5, M107 and M71 which enables a direct comparison. M107 
and M71 do not show any anticorrelation in either of the studies. 
\citet{carretta02} only have upper limits for Al and Mg in M15, while we were able to make direct measurements, 
but even with upper limits the two studies show similar anticorrelations with 
largest spreads in both Al and Mg from the whole sample of clusters. The largest difference between the two studies 
can be seen in case of M5, where our values span a range of $-$0.3~dex to +1.1~dex, while the Al abundances of \citet{carretta02} 
are limited to between $-$0.2 and 0.7~dex. The difference may be explained by 
a simple selection effect, as \citet{carretta02} observed 13 stars in M5, while our sample size is 122 and covers almost the full 
extent of the RGB. 

We discuss the extent of anticorrelation through differences in the average of the Mg abundance between the populations, 
divided by the estimated uncertainty in this average Mg difference. We use the final combined errors from Table 4. 
Table 8 lists the errors of the average differences with values of how much $\sigma$ 
detection is each difference. We find that in M15 (4.3$\sigma$), M92 (9.7$\sigma$), M13 (8.6$\sigma$), and M3 (7.3$\sigma$) 
the differences are statistically significant and the Mg-Al anticorrelation exists. Two clusters, M53 (3.7$\sigma$) and M5 (3.8$\sigma$), 
also show large $\sigma$ detections, however, we would like to use a more conservative approach to the detection of 
anticorrelations out of a concern that our errors may be underestimated, given the degeneracies between stellar 
parameters and abundances in our analysis. M2 stands out as having no statistically significant Mg-Al anticorrelation. 

The summed abundance A(Mg+Al) is expected to be constant as a function of T$_{\rm eff}$ when material is 
completely processed through the Mg-Al cycle, and 
that is what our results show in Figure \ref{fig:mgaladd}. 
The only exception is M107, but the slight correlation is due Mg correlating with temperature (see Figure \ref{fig:femgal}), 
and this trend was not removed here.

Al is expected to correlate with elements enhanced by proton-capture reactions
(N, Na; Figure \ref{fig:sicor}) and anti-correlate with those depleted in H-burning at high temperature (O, Mg; Figure \ref{fig:sicor}). 
The Al-O anticorrelations and Mg-O correlations can be clearly seen in our data in Figure \ref{fig:sicor} for all clusters except M53, M107 
and M71. The small number of stars observed in NGC~5466 does not allow us to investigate the correlations in that cluster in detail. 
However, the slight correlation of Al with Si that can be seen in M15 in Fig. \ref{fig:sicor} is the evidence of $^{28}$Si 
leaking from the Mg-Al cycle, which is an intriguing result. Si participating in the light-element abundance pattern was first 
reported in NGC~6752 by \citet{yong05}. Since then, \citet{carretta02} have also found Si enhancement correlated with Al  
in NGC~2808. Those authors' interpretation was that it is only in low-metallilcity clusters, where the AGB stars burn slightly hotter, 
or in high-mass clusters, where the chemical enrichment is more efficient, that a Si enhancement will be observed in 
second-generation GC stars. The difference in Si abundance between the two XD-identified populations, listed in in Table 8, 
is significant relative to the error on that measurement. The Si-Al correlation in M15 is also accompanied by a Si-Mg 
anticorrelation (Figure \ref{fig:sicor}), which is further evidence of $^{28}$Si being produced by hot bottom burning in AGB 
stars \citep{karakas01}.

\subsection{The spread of Al abundances}

The spread of Al abundances (Figure \ref{fig:rms}) also 
increases significantly below [Fe/H]=$-$1.1, thus we can conclude that high mass, low metallicity AGB polluters that 
are able to process material at high, 60$-$70 million K, play an increasingly larger role in more metal-poor cluster. 
This behaviour of Al abundances was previously observed by \citet{carretta02}, but the larger range of Al values 
presented in this paper make this correlation clearer.

Theoretical AGB  nucleosynthesis modeling indeed predict this behavior \citep{ventura01, ventura02}. 
The high-mass AGB stars reach higher temperatures at the bottom of the 
convective envelope; i.e., stronger HBB and advanced (Mg-Al) nucleosynthesis occurs 
with decreasing metallicity. We are seeing very advanced Mg-Al nucleosynthesis 
in the most metal poor clusters such as M15 and M92, while the most metal-rich 
clusters like M107 and M71 do not show this, as theoretically expected, if the 
high-mass AGB stars are the polluters. Also, this is corroborated by the Al-O 
anticorrelation; the most Al-rich stars are O-poor showing the effects of 
very strong hot bottom burning (HBB), because HBB proceeds completely and destroys O.

An other related issue is that HBB is activated for lower masses with decreasing 
metallicity. High-mass and very low-metallicity 
AGB stars don not exists in GCs today due to their extremely short lifetimes, but this trend seems to be confirmed both theoretically 
\citep{ventura01, ventura02} and observationally at least for solar 
metallicities down to that of the Magellanic Clouds. From the observational point 
of view, we know that high-mass AGB stars in the Small Magellanic Cloud 
([Fe/H]=$-$0.7) activate HBB for progenitor masses (M $>$3 M$_{\odot}$) lower 
than their solar metallicity counterparts (M $>$ 4 M$_{\odot}$) 
\citep{garcia02, garcia01}. Thus, at the lowest metallicities in M15 and M92 we also would 
expect more HBB AGB stars (i.e., with several degrees of HBB and Mg-Al nucleosynthesis), 
because the minimum stellar mass to activate the HBB process (and advanced Mg-Al nucleosynthesis) 
decreases with decreasing metallicity. Thus, we conclude that more polluted material would 
be present at the lowest 
metallicities.

Because of these reasons, we believe that our results support and add some evidence 
to the high-mass AGBs as GCs polluters.

\subsection{C-N}

\begin{figure}[!ht]
\centering
\includegraphics[width=3.5in,angle=0]{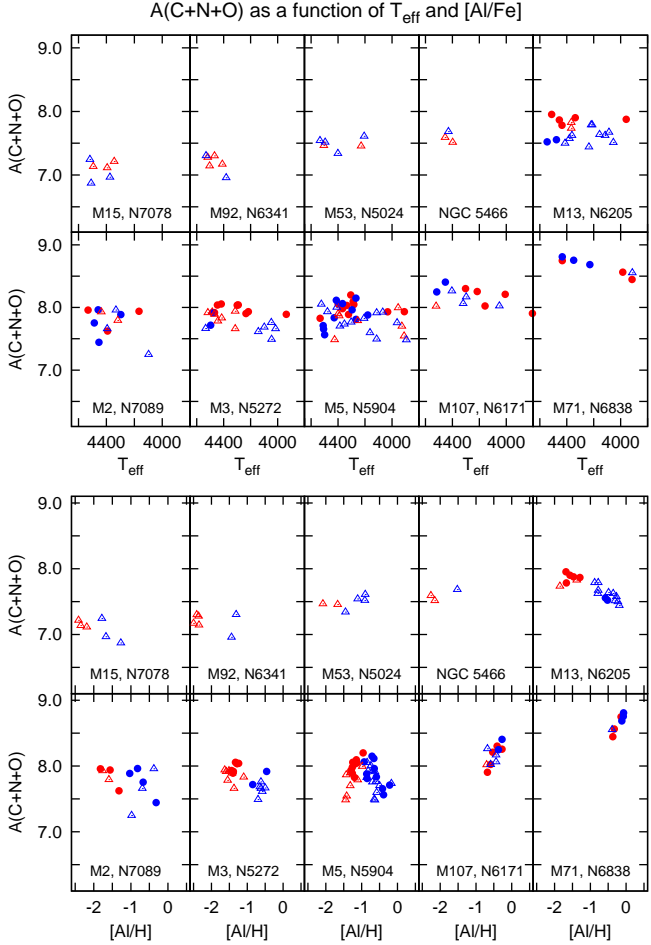}
\caption{Upper panels: The sum of C, N and O as a function of effective temperature. Upper limits are denoted by open triangles. 
For explanation of color coding, please see description of Figure 8. Lower panels: The sum of C, N and O as a function of [Al/H].
A clear correlation is visible in M13, M2, M3, and M5, 
for more discussion see Section 6.4. 
}
\label{fig:cnoadd}
\end{figure}

Carbon, nitrogen and sometimes oxygen are influenced by two independent processes in GC giants: 
a primordial anticorrelation, with the same first-generation sources as the O-Na and Mg-Al patterns; 
and also stellar evolution, driven by circulation between the hydrogen-burning shell and the 
surface \citep{sweigart01, angelou01}. As discussed in 
the previous section, we see large star-to-star variations in N abundance in all clusters; however, our inability to measure C 
abundances below [Fe/H]=-1.7~dex leads us to only having upper limits, and therefore we restrict the 
discussion of C-N anticorrelations to the more metal-rich clusters. 

Figure \ref{fig:cnanti} shows [N/Fe] versus [C/Fe] for all clusters. Deep mixing has to be taken into account when 
discussing the CN anticorrelations, so in M13, M2, M3, and M5, where we see deep mixing clearly, the correlation of [C/Fe] 
with T$_{\rm eff}$ was removed. Because deep mixing is not visible in M107 and M71, we used the original values. 
A clear anticorrelation cannot be 
seen in any of the clusters. While M107 and M71 show weak anticorrelations, neither of these is 
statistically significant. The anticorrelation itself may be obscured by the relatively large errors 
for [C/Fe] and [N/Fe]. As mentioned in Section 5.1, we see correlations of [C/Fe] with temperature in M13, M3, M2, and M5, but 
these are not accompanied by increasing [N/Fe] at the same time. We believe that this is due to our inability to 
measure C and N accurately: most of our abundance determinations are upper limits or close to them. If the 
anticorrelations exist, they must span ranges smaller than our uncertainties, which are always higher than 
0.14~dex in [C/Fe] and 0.12~dex in [N/Fe].
It is interesting to note that in M3, M107 and M71 there is a clear CN weak and CN strong group of stars, 
while in all more metal poor clusters, N abundances fill out a broad distribution.

\subsection{C+N+O}

We have C, N, and O abundances available for a large number of stars, thus we are able to investigate the 
C+N+O content in each cluster in detail. Studies from the literature showed that the C+N+O content in globular 
clusters are fairly constant to within 0.3~dex \citep{ivans02, carretta04, smith03}, except for N1851 where 
\citet{yong04} find a spread of 0.57~dex. According to \citet{yong04} the large spread in 
C+N+O in N1851 is probably attributed to larger than usual pollution from lower mass AGB stars than in other 
clusters, but these results were questioned by \citet{villanova01} as they did not see spread of 
C+N+O larger than 0.3~dex.

We also see near constant C+N+O in our sample, which is consistent with
the material in these stars having undergone CNO cycling in
the first-generation of stars in the RGB phase. In Figure \ref{fig:cnoadd} we show A(C+N+O) as a function of T$_{\rm eff}$. 
The spread is significantly larger than what is reported in literature, between 0.4 and 0.6~dex 
for M13, M2, M3, and M5. The spread in these clusters is at the level of what has been found in N1851, but we think 
this is mostly associated with large uncertainties of [N/Fe] measurements. We do not find clear 
separation in the amount of A(C+N+O) between the two populations found in any of the clusters. 

As previously mentioned, there are currently two leading models to explain the nature of the polluters: the first assumes that 
intermediate-mass AGB stars that are thermally pulsing and undergoing hot bottom 
burning expel material to the intra-cluster medium by strong mass loss \citep{ventura01}, while the second 
assumes that the pollution comes from hot, fast rotating stars \citep{decressin01}. 
According to the first theory, intermediate-mass HBB-AGB stars could produce or not produce large 
C+N+O variations \citep[see, e.g.,][and references therein]{yong04}. This is because the C+N+O predictions (which basically 
depend on the number of third dredge-up episodes) for these stars are extremely dependent on the theoretical modelling, 
the convective model and the mass loss prescription used \citep[see, e.g.,][]{dantona01, karakas02}. 
A (C+N+O)-Al correlation is not clearly visible in the lower panel of Figure 10.

According to the first theory, AGB stars may explain a large spread in A(C+N+O) because they are expected to produce a 
substantial increase in the C+N+O abundance as they enhance Na and Al and deplete O and Mg \citep{yong04}. This should 
result in a (C+N+O)-Al correlation, which is not clearly visible in the lower panel of Figure \ref{fig:cnoadd}. Weak correlations 
in M107 and M71, and weak anticorrelations in M13 and M3 are visible, 
but they are probably the result of large uncertainties in C and N. More accurate measurements of [N/Fe] will help to clarify the 
possible (C+N+O)-Al correlation.

\section{Summary}

We investigated the abundances of nine elements for 428 stars in ten globular clusters using APOGEE DR10 spectra. 
A homogeneous analysis of these GCs has not been accomplished previously, something that APOGEE is uniquely able 
to do because it can observe all the bright GCs in the northern hemisphere. 
A semi automated code called \textit{autosynth} was developed to provide abundances independent of those derived by ASPCAP. 
Our main goal was to examine the stellar populations in each cluster using a homogeneous dataset. Based on our 
abundances, we find the following:

1. From the examination of the star-to-star scatter in $\alpha-$element abundance, only O and Mg show an abundance scatter 
larger than our measurement errors. However, this is expected, because O and Mg are part of the well-known 
globular cluster light-element abundance anticorrelations.

2. Population analysis using [N/Fe] for M107 and M71 and [Al/Fe] in the rest of the clusters confirms 
that each cluster can be divided into two discrete populations, though in M107 and M71 that 
division is clearest in the N-Al plane, whereas in the other clusters it is best made in the Mg-Al plane.

3. The anticorrelation of Mg and Al is clearly shown in M15, M92, M13 and M3, and to a lesser extent in M53 and M5. 
Interestingly, M2, which is also a metal poor cluster, does not have a Mg-Al anticorrelation. The increased number of 
stars in our sample compared to the literature enables us to discover more Al-rich stars, making the 
anticorrelation clearer in this dataset than in previous studies.

4. The spread of Al abundances increases with decreasing cluster metallicity. This is in agreement with theoretical
AGB nucleosynthesis predictions by \citet{ventura01, ventura02}. This suggests that high-mass HBB AGB stars, which are able to
enrich Al while destroying O, are more common polluters in metal-poor clusters than in metal-rich ones.

5. A correlation between Al and Si in M15 indicates particularly high-temperature hydrogen 
burning in the stars that contributed the abundance pattern of the second-generation stars. This is 
consistent with Si enhancement in globular cluster stars as discussed in the literature, where it 
appears to be confined to low-metallicity and high-mass clusters.

6. Besides accessing northern GCs, APOGEE is also unique among other spectroscopic surveys, because it is capable of 
measuring abundances of CNO. The total abundance of C+N+O is found to be constant in each cluster, within our errors. 
Our data set do not show an unambiguous correlation between A(C+N+O) and [Al/H]. In principle, this finding is not 
against the idea of intermediate-mass HBB-AGB stars as the primary source of chemical self-enrichment in globular
clusters. However, more precise measurements of C and N abundances will clarify this issue, and we expect that 
ASPCAP will be able to provide these in the future.


\acknowledgements{Szabolcs Meszaros is especially grateful for the technical expertise and 
assistance provided by the Department of Astronomy at Indiana University.
We thank Eileen D. Friel, Catherine A. Pilachowski and Enrico Vesperini for their detailed comments 
during the development of autosynth. 

Sarah Martell acknowledges the support of the Australian
Research Council through DECRA Fellowship DE140100598. Sara Lucatello acknowledges partial support from grant 
PRIN MIUR 2010-2011 “Chemical and dynamical evolution of the Milky Way and Local Group galaxies”. 
Jo Bovy was supported by NASA through Hubble Fellowship 
grant HST- HF-51285.01 from the Space Telescope Science Institute, which is operated by the Association of 
Universities for Research in Astronomy, Incorporated, under NASA contract NAS5-26555. Timothy C. Beers acknowledges 
partial support for this work from grants PHY 08-22648; Physics Frontier Center/{}Joint Institute or Nuclear
Astrophysics (JINA), and PHY 14-30152; Physics Frontier Center/{}JINA
Center for the Evolution of the Elements (JINA-CEE), awarded by the US
National Science Foundation. 
Katia Cunha acknowledges support for this research from the National Science Foundation (AST-0907873). 
Verne V. Smith acknowledges partial support for this research from the National Science 
Foundation (AST-1109888). D.~A.~Garc\'{\i}a-Hern{\'a}ndez and Olga Zamora acknowledge support provided by the Spanish Ministry of Economy 
and Competitiveness under grant AYA-2011-27754.

Funding for SDSS-III has been provided by the Alfred P. Sloan Foundation, the Participating Institutions, the 
National Science Foundation, and the U.S. Department of Energy Office of Science. The SDSS-III web site is 
http://www.sdss3.org/.

SDSS-III is managed by the Astrophysical Research Consortium for the Participating Institutions of the SDSS-III 
Collaboration including the University of Arizona, the Brazilian Participation Group, Brookhaven National Laboratory, 
University of Cambridge, Carnegie Mellon University, University of Florida, the French Participation Group, the 
German Participation Group, Harvard University, the Instituto de Astrofisica de Canarias, the 
Michigan State/Notre Dame/JINA Participation Group, Johns Hopkins University, Lawrence Berkeley National Laboratory, 
Max Planck Institute for Astrophysics, New Mexico State University, New York University, Ohio State University, 
Pennsylvania State University, University of Portsmouth, Princeton University, the Spanish Participation Group, 
University of Tokyo, University of Utah, Vanderbilt University, University of Virginia, University of Washington, 
and Yale University. 

}

\end{document}